\documentclass[]{article}

\pdfoutput=1 

\usepackage{amssymb}
\usepackage{amsmath,amsfonts}
\usepackage{algorithm}
\usepackage{algorithmic}
\usepackage{url}
\usepackage{cleveref}
\usepackage{eurosym}
\usepackage{todonotes}
\usepackage{xcolor}
\usepackage{pgfplots}
\pgfplotsset{compat = newest}
\usepackage{comment}
\usepackage{tikz}
\usepackage{authblk}
\usepackage[backend=bibtex]{biblatex} 
\title{Threshold-Based Algorithms for an Online Rolling Horizon Framework Under Uncertainty - With an Application to Energy Management}
\author[1]{Jens H\"onen}
\author[1]{Johann L. Hurink}
\author[2,3]{Bert Zwart}
\affil[1]{Faculty of EEMCS, University of Twente, Enschede, The Netherlands}
\affil[2]{Department of Mathematics and Computer Science, Eindhoven University of Technology, Eindhoven, The Netherlands}
\affil[3]{Centrum Wiskunde \& Informatica (CWI), Amsterdam, The Netherlands}
\date{}

\begin{document}

\maketitle

\begin{abstract}
Decision problems encountered in practice often possess a highly dynamic and uncertain nature. In particular fast changing forecasts for parameters (e.g., photovoltaic generation forecasts in the context of energy management) pose large challenges for the classical rolling horizon framework. Within this work, we propose an online scheduling algorithm for a rolling horizon framework, which directly uses short-term forecasts and observations of the uncertainty. The online scheduling algorithm is based on insights and results from combinatorial online optimization problems and makes use of key properties of robust optimization. Applied within a robust energy management approach, we show that the online scheduling algorithm is able to reduce the total electricity costs within a local microgrid by more than 85\% compared to a classical rolling horizon framework and by more than 50\% compared to a tailor-made dynamic, yet still offline rolling horizon framework. A detailed analysis provides insights into the working of the online scheduling algorithm under different underlying forecast error distributions.
\end{abstract}

\section{Introduction}\label{S1Intro}
Solving real-world optimization problems usually involves a large variety of constraints, representing different aspects of the considered setting, such as legal regulations, device properties, or physical laws. In addition to the often highly complex and influencing constraints, large parts of the model parameters are subject to uncertainty and are therefore not known perfectly in advance. This uncertainty aspect requires particular solutions techniques, such as e.g., robust optimization, stochastic programming, or the integration into a rolling horizon framework. Due to improved forecasts over time or new measurements, this uncertainty however may reduce over time, leading to dynamic optimization problems under uncertainty. Such problems appear in many areas, from energy management, \cite{RollForecast2024Ghadimietal}, over production planning and scheduling, \cite{ProdPlanning2022Ghadimietal}, to logistics, \cite{LogisticsRH2012Nielsen}. What most of these problems have in common is that decisions need to be made over time. As such, a decision taken now may further restrict what can be done in future decision steps. Therefore, the impact of decisions should not only be considered for the given solution known at the current time (slot) but also some form of look-ahead into the future needs to be considered. To summarize, the past, the present and the future all impact the decision-making process, resulting in a highly dynamic optimization problem under uncertainty.\\

Within this work, we consider a dynamic optimization problem from the energy domain as the primary example. It concerns a (joint) energy management approach for a microgrid. Here, the microgrid is a set of households connected via the electricity grid (e.g., a neighborhood), which has to make a variety of time-critical decisions, such as charging and discharging batteries or electric vehicles (EVs) or placing bids on markets. These decisions have to be made in a setting with a high degree of uncertainty.\\

Within the last years, quite some progress has been made within the area of energy management or energy trading approaches (see e.g., \cite{ClassificationLET2023Hoenenetal}, \cite{SurveyAGTlocalEnergyTrading2019PilzAlFagih} \cite{SurveyMGEMS2020Elmouatamidetal}, \cite{SurveyEnergyManagement2019Garciaetal}, \cite{SurveyMicrogridEMS2018Ziaetal} for recent survey papers and literature studies). Throughout this research, different techniques, ranging from rolling or receding horizon to adaptive robust optimization or stochastic programming have been successfully applied to tackle such (large-scale) dynamic energy management problems under uncertainty. The advantage of the rolling horizon approaches (see e.g., \cite{RHMGsmoothing2022Lietal}, \cite{RHMGcorrectionfeedback2018Tahaetal}, \cite{RHMGChile2013Palmaetal}, \cite{RHMG2Stage2018Elkazaz}, \cite{RHMGEMconvex2020Elkazazetal}, \cite{RHsimMGEM2015Silventeetal}, \cite{RHisolatedMG2014Olivaresetal}) lies within the repeating solving of subproblems of the original problem. In each solving process, which we denote as an iteration of the rolling horizon, updated forecasts and new information of uncertain data may be used to improve the resulting solution. Mathematical techniques, such as robust optimization or stochastic programming on the other hand directly focus on the feasibility of the resulting solution by using additional information of the uncertain data, such as underlying probability distributions or uncertainty sets. These techniques have become quite popular for applications within the energy domain (see e.g., \cite{AROUnitCommitment2013Bertsimasetal}, \cite{AROlongTermScheduling2020Moretti}, \cite{AROisolatedMG2021Sadeketal}, \cite{AROuncertainArrivalEV2018ChoiHussainKim}, \cite{ROV2G2015BaiQiao}, \cite{RobustEMMG2022Hoenenetal}, \cite{ROEnergyDispatchDA2015Zugnoetal}, \cite{RODAhydroelectric2023Zhongetal}) as they match well with the feasibility focus of the current energy system. One way how to combine the advantages of robust optimization and the rolling horizon is to use robust or stochastic techniques as solution techniques for the iterations of a rolling horizon (see e.g., \cite{RHaaroSeasonalStorage2020Castellietal}, \cite{RHdynamicEM2022Yodasetal}, \cite{RHstochasticEM2018Silventeetal}, \cite{AROisolatedMG2019Laraetal}, \cite{ROMGensembeWind2017Craparoetal}, \cite{RHrobustLEM2022Nikkhahetal}, \cite{AROmobileEES2022Luetal}, \cite{ROsizingESdynamicUS2022KeyvandarianSaif}, \cite{ARODynWind2015LorcaSun}, \cite{ROUnitCommDynamicWind2017LorcaSun}, \cite{RHhydrothermal2012Guigesetal}, \cite{RHLoadSchedding2018Devineetal}). This allows to use updated (real-time) information on uncertain data to incorporate the uncertainty and reduce stochasticity.\\

One possible disadvantage of the classical rolling horizon framework is the use of equidistant spacing between iterations, which often does not fit well with the uncertain nature of parameters. Photovoltaic (PV) generation, for example, is assumed to be of major importance for the energy transition (see e.g., \cite{PVcapacityEU2021Kougiasetal}), but also difficult to forecast. As such, it is an uncertain parameter within an energy management problem, but obviously, its uncertainty is not spread equally among the day but is limited to daytime. Therefore, the iterations of a classical rolling horizon framework, which are scheduled during the night, do not contribute to an improvement of the solution as no new short-term PV forecasts are available. To be able to better capture the improvements in PV forecasts, in earlier work \cite{DynRHRMES2023Hoenenetal} we proposed a dynamic rolling horizon framework. The key idea of this dynamic rolling horizon framework is to go beyond the fixed time distance between consecutive iterations and schedule them in a more flexible way to better capture the updated PV forecasts. For this, we developed a dynamic (offline) scheduling tool, which identifies good starting time slots for the iterations of a rolling horizon. This scheduling tool is based on a version of a knapsack problem and it uses the structural knowledge of the underlying uncertainty sets from robust optimization.\\

Although the combination of the dynamic scheduling tool with a rolling horizon yields large improvements over the classical rolling horizon, some weaknesses have been identified. Firstly, the knapsack problem determines starting time slots based on average realizations of future uncertainty. This may be a reasonable choice if the expectation of the uncertainty realization is close to the average. However, for a skewed underlying uncertainty distribution, this could considerably affect the outcome of the optimized starting time slots. The second disadvantage follows from the offline nature of the decision-making process and is even more important. The decision when to start the iterations of the rolling horizon is already made at the start of the considered time horizon, and therefore there is no option to react to particularly good (or bad) forecasts or observations of uncertainty. Specifically, given the nature of the considered problem, it is of high importance to be able to react to extreme or unusual events.\\

Based on the above considerations, within this work, we design an online version of the dynamic rolling horizon framework, which improves upon both of the above-mentioned disadvantages. Instead of deciding upon the starting time slots in advance, the online rolling horizon framework decides on the fly whether to start an iteration or not. Thereby, it is able to react to updated forecasts or observed realizations of uncertainty and can make better decisions regarding the starting of iterations. In a first step, we generalize the knapsack formulation, first introduced in \cite{DynRHRMES2023Hoenenetal}, beyond the explicit application of an energy management problem. We then reduce the generalized formulation to an instant of the $k$-edge longest path problem in directed acyclic graphs, which we show can be solved efficiently via a dynamic programming approach. The outcome of this path problem just provides the threshold for the online threshold algorithm, which is run for each time slot to decide whether or not to start an iteration of the rolling horizon. The online threshold algorithm is inspired by algorithms with a constant competitive ratio for various online problems, such as the online knapsack, multi-secretary, or prophet inequality problem. In addition, it makes use of the main principles of robust optimization and is thereby able to react to extremely good or bad forecasts or realizations of uncertainty. Within a case study, we show that the online rolling horizon framework is able to improve the solution by over 85\% compared to the classical rolling horizon approach with a fixed step size, and by over 50\% compared to a tailor-made offline rolling horizon scheme, as presented in \cite{DynRHRMES2023Hoenenetal}. To sum up, the main contributions of this work are:
\begin{itemize}
	\item We propose a novel online rolling horizon framework, which can adapt the starting time slots of its iterations to updated forecasts and observed uncertainty. 
	\item We derive an efficient algorithm for solving the $k$-edge longest path problem subject to the constraint that a given subset of the nodes has to be part of the path.
	\item We conduct a detailed case study, which shows the potential of the online rolling horizon framework compared to various offline rolling horizon schemes.
\end{itemize}

This work is an extension of earlier work on dynamic scheduling of starting time slots of a rolling horizon in the context of a local energy management problem under uncertainty, \cite{DynRHRMES2023Hoenenetal}. The considered underlying setting is therefore the same, but both, the used techniques to design the online framework, as well as the analysis, have not been considered before. In addition, a refined analysis of the knapsack problems as well as the underlying error distributions is presented.\\

The paper is structured as follows: The central mathematical optimization model is derived and presented in Section \ref{S2Model}, while Section \ref{S3OnlineAlg} introduces the different online scheduling algorithms for the rolling horizon framework. Section \ref{S4ApplicationScenario} presents the application scenario of a robust energy management approach and in Section \ref{S5Analysis}, we present, analyze and discuss the results of a case study. We conclude the work in Section \ref{S6Conclusion} and provide interesting research directions for future work.

\section{Model}\label{S2Model}
In this section, we introduce the essential mathematical components of the online algorithms, presented in Section \ref{S3OnlineAlg}. Inspired by the results of the knapsack approach in \cite{DynRHRMES2023Hoenenetal} to identify good starting time slots for the iterations of a rolling horizon, we first generalize the knapsack model beyond the specific application of a robust energy management approach. In a second step, we reduce the generalized knapsack problem into a longest path problem and provide insights into the computational complexity and running time. 

\subsection{Input Parameter and Notation}
To decide when to execute iterations of the rolling horizon, we consider the following input data:
\begin{itemize}
	\item In applications of a rolling horizon framework, the considered time horizon of the problem is usually split up into a set of equidistant time slots. $\mathcal{T} = \left\lbrace 0,1,\ldots,T \right\rbrace$ denotes the set of time slots. As we consider an ongoing rolling horizon, we assume that an initial solution is given, which could be the solution of the last iteration of the previous time horizon. Within $\mathcal{T}$, let time slot $0$ represent the time slot of this initial solution.
	\item $c \in \mathbb{R}_{\geq 0}^{\vert\mathcal{T}\vert\times \vert\mathcal{T}\vert}$ represents the information gain over time within the rolling horizon, whereby the parameter $c_{s,t}$ denotes the information gained when starting an iteration of the rolling horizon at time slot $t$, given that the previous iteration started at time slot $s$.
	\item $k \in \mathbb{N}_{>0}$ is a positive integer denoting the maximal number of iterations of the rolling horizon over the time horizon $\mathcal{T}$. This upper bound on the number of iterations may be imposed by various aspects, such as e.g., computational resources, limited (desired) communication, or other setting-dependent constraints.
	\item $\mathcal{M} = \left\lbrace i_0,i_1,\ldots,i_m \right\rbrace \subset \mathcal{T}$ is a given set of time slots, which needs to be part of the selected time slots, where an iteration of the rolling horizon is executed. As time slot $0$ corresponds to the initial solution, we assume that $0=i_0 \in \mathcal{M}$. In general, $\mathcal{M}$ could represent certain time slots, in which decisions need to be communicated or submitted to markets or other internal or external entities within the considered setting.
\end{itemize}

\begin{figure}[ht]
	\centering
	\includegraphics[width=10.5cm]{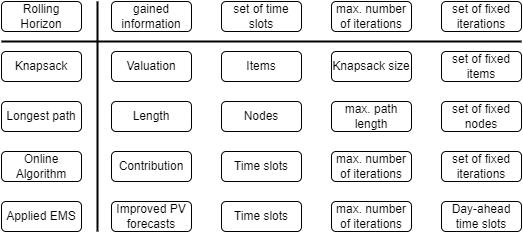}
	\caption{Overview and mapping between the used notation within the different models.}
	\label{FigMapping}	
\end{figure}

Within the following, we introduce two different models, which are essential components of the online scheduling algorithms presented in Section \ref{S3OnlineAlg}. As these models are based on well-known combinatorial optimization problems, we stick to their usual notation. Figure \ref{FigMapping} provides a mapping between the notation of the various models and the online scheduling algorithm.

\subsection{Knapsack Formulation}\label{S2.1Knapsack}
As for the knapsack model in \cite{DynRHRMES2023Hoenenetal}, the items of the knapsack correspond to the potential starting time slots of iterations of the rolling horizon, while the valuation of an item corresponds to the additional information it can provide when starting an iteration in this time slot. In our setting, all items have a weight of one, while the knapsack size corresponds to the maximal number of iterations during the time horizon. This also implies that we have a topological ordering on the items of the knapsack problem, which directly follows from the structure of the time slots.\\

Let $x_{t}$, $t \in \mathcal{T}$, be a binary variable indicating whether to include item $t$ into the knapsack or not. As the valuation of an item $t$ depends on the choice of its predecessor $s$, let $y_{s,t}$, $s<t$, be an auxiliary binary variable to link the chosen items to their corresponding valuation. The objective is to maximize the value of the chosen items:
\begin{equation}\label{mknapBaseobj}
	\max \sum_{s,t \in \mathcal{T}} c_{s,t}y_{s,t}.
\end{equation}
The classical knapsack constraint, which limits the number of items to at most $k$, is
\begin{equation}\label{mknapBaseeq1}
	\sum_{t \in \mathcal{T}}x_t \leq k+1.
\end{equation}
Note that we use an upper limit of $k+1$, as item $0$ is also included, but should not count as an online iteration. To ensure that the items in $\mathcal{M}$ are chosen, we add constraints
\begin{equation}
	x_t = 1 \quad \forall t \in \mathcal{M}.
\end{equation}
The following constraints link the variables $x_t$ with the auxiliary variables $y_{s,t}$ and thereby establish a connection to the valuations:
\begin{equation}
	y_{s,t} \leq x_t \quad \forall s,t \in \mathcal{T},
\end{equation}
\begin{equation}
	y_{s,t} \leq x_s \quad \forall s,t \in \mathcal{T}.
\end{equation}
To ensure that only the valuations between neighboring items are counted, we ensure that each item counts at most once as the beginning and at most once as the finishing item within the $y_{s,t}$ variables, which have a value of 1:
\begin{equation}\label{mknapBaseeq4}
	\sum_{s \in \mathcal{T}}y_{s,t} \leq 1 \quad \forall t \in \mathcal{T},
\end{equation}
\begin{equation}\label{mknapBaseeq5}
	\sum_{t \in \mathcal{T}}y_{s,t} \leq 1 \quad \forall t \in \mathcal{T}.
\end{equation}
To ensure that only the valuations between neighboring items are counted, we add the following constraints:
\begin{equation}\label{mknapBaseeq6}
	1 - x_t + y_{s,t} \geq y_{s,l} \quad \forall s,t,l \in \mathcal{T}, s<t<l.
\end{equation}
If items $s<t<l$ are all taken, then $x_t=1$. The above constraints then simplify to $y_{s,t} \geq y_{s,l}$, which together with constraint (\ref{mknapBaseeq5}) ensures that $y_{s,l}=0$ and the valuation $c_{s,l}$ does not add to the objective. If $s$ is not taken, then all variables $y_{s,.}$ are automatically zero and if $t$ is not taken ($x_t = y_{s,t} = 0$), then the constraint simplifies to $1 \geq y_{s,l}$, which does not lead to any restrictions. Hence, the objective function (\ref{mknapBaseobj}), together with constraints (\ref{mknapBaseeq1})-(\ref{mknapBaseeq6}) forms the knapsack problem.

\subsection{Reduction to Graph-Based Problem}\label{S2.2Path}
Analyzing constraints (\ref{mknapBaseeq4})-(\ref{mknapBaseeq6}), it becomes obvious that the $y$ variables actually have to form a path on some graph over $\mathcal{T}$. Constraints (\ref{mknapBaseeq4}) and (\ref{mknapBaseeq5}) ensure that there is at most one in-going and at most one out-going arc for each pair $s,t \in \mathcal{T}$, while equation (\ref{mknapBaseeq6}) ensures that only arcs between consecutive time slots may be taken. Coupled with the objective (\ref{mknapBaseobj}) and the non-negativity of the $c$ values, an optimal solution corresponds to a path of at most $k$ arcs. Summarizing, the above knapsack problem can be reduced to a longest path problem with at most $k$ arcs, visiting all nodes in $\mathcal{M}$, on a directed graph, spanned on the set of time slots $\mathcal{T}$.\\

To formalize, the  graph is a complete directed acyclic graph (DAG) $\mathcal{G}=(\mathcal{V}, \mathcal{A}, c)$, in which the node set $\mathcal{V}$ corresponds to the time horizon $\mathcal{T}$, enlarged by an artificial node $T+1$. The set of directed arcs $\mathcal{A}$ consists of arcs $(s,t)$ for all pairs $s,t \in \mathcal{V}$, $s<t$, and the arc length $c$ is given by the information gains $c$. Furthermore, the natural ordering of the time slots corresponds to the unique topological ordering of the nodes of the graph. For completeness, the arc length for the arcs going into the artificial node $T+1$ is given by $c_{s, T+1}=0$. The knapsack problem of finding an optimal solution with at most $k$ items, using all items in $\mathcal{M}$ now translates into the problem of finding a $0-(T+1)$-path of at most $k+1$ arcs, which visits all nodes in $\mathcal{M}$ and maximizes the length along this path. \\

Due to the topological ordering of the DAG $\mathcal{G}$, this problem can be solved using two different \textit{dynamic programming} approaches in $\mathcal{O}(T^2k)$ time. In a first step, we compute for $j=0, \ldots,m$ the longest paths between nodes $i_j, i_{j+1} \in \mathcal{M}$ with exactly $h \in \{1, \ldots, \min(k+1,i_{j+1}-i_j)\}$ arcs, where $i_{m+1}=T+1$. This can be done in time $\mathcal{O}(T_i^2k)$ for each pair $i_j$ and $i_{j+1}$, whereby $T_i=i_{j+1}-i_j$. Summing up over all pair of nodes $i_j, i_{j+1}$, we have a total complexity of $\sum \mathcal{O}(T_i^2k) \leq \mathcal{O}(T^2k)$, as $\sum T_i = T$. In a second step, we reduce the original graph to the node set $\mathcal{M}\cup\left\lbrace T+1 \right\rbrace$ and use the previously computed longest paths with $h$ arcs between neighboring nodes. By a second dynamic programming approach, the longest path from node $0$ to node $T+1$ using at most $k+1$ arcs can be calculated in a straightforward way in an iterative fashion. The running time of this second dynamic program is $\mathcal{O}(mk^2)$, leading to an overall complexity of $\mathcal{O}(T^2k)$, as $T>k>m$.

\section{Online Algorithm}\label{S3OnlineAlg}
In this section, we first review the literature on similar online optimization problems and then introduce and derive the main principles and algorithms for the online scheduling of the rolling horizon.

\subsection{Online Literature}
As mentioned in Section \ref{S1Intro}, an online approach to schedule the starting time slots of a rolling horizon may have several benefits over an offline scheduling approach. To gain insights into the working of online algorithms, a review of solution approaches for various online combinatorial optimization problems is given, which are based on a similar online problem structure as the online scheduling of the iterations of a rolling horizon in this study. Note that the reviewed literature analyzes the online problems from a theoretical perspective, where the main goals are to derive (constant) competitive online algorithms. Although this theoretical focus differs from the rather practical application in our study, high-level concepts and ideas can still be applied to our problem. \\

Online optimization has been an important research direction in the area of (combinatorial) optimization for many years now. Online problems, in which items arrive one by one, and for each item it has to be decided irrevocably whether to keep or discard it, can be found in online versions of the knapsack problem, the (multi-)secretary problem, \cite{SecProbOpt1963Dynkin}, resource assignment problems, \cite{SeqAP2006Chunetal}, or prophet inequality problems, \cite{CombiProphetInequ2017RubinsteinSingla}. Next to some negative results on the competitiveness of online algorithms for some of these problems \cite{StochKnapsack1995Marchettietal}, promising online algorithms have been developed under some (mild) assumptions. These assumptions range from knowledge of distributions or ranges of values for certain item parameters (\cite{StochKnapsack1995Marchettietal}, \cite{OKPKeywordAuctions2008Zhou}, \cite{OnlineKnapsackDeparture2022Sunetal}), to the order in which items may arrive (\cite{ProphetSecretaryStaticThreshold2022ArnostiMa}, \cite{OGAPHistory2023Liuetal}). Independent of these different assumptions, many of the proposed competitive online algorithms are based on the same high-level solution approach with small differences in specific elements of the algorithms, often reflecting the various assumptions.\\

In \cite{ProphetSecretaryStaticThreshold2022ArnostiMa}, the values of the items are drawn independently from known distributions and the items arrive in a uniformly random order. The focus is to derive a static threshold, such that the values of the first $k$ items with values exceeding this threshold are maximized in expectation. The decision rule is therefore based on two important parameters, namely the value of the items, as well as the static threshold. \\

In \cite{OMKPCompetetive2021Yangetal}, the used threshold is not static, but a linearly increasing function of the utilization rate of the knapsack (in multiple dimensions). In addition, the weight of the item is also included in the threshold. Again, the value of an item gets known as soon as the item arrives and an irrevocable decision regarding the usage of this item has to be made. Furthermore, a second online algorithm is proposed, which admits an exponential threshold function. This threshold function is again increasing in the utilization rate of the knapsack and also includes the weight as a factor. In comparison to the linear threshold function, this exponential threshold is more conservative in admitting items and prefers to reserve available space for future high-value items. Further exponential threshold functions depending on the utilization of the knapsack are presented in \cite{OKPKeywordAuctions2008Zhou}, \cite{OnlineKnapsackDeparture2022Sunetal}, \cite{OnlineCloudPrices2017Zhangetal}, and \cite{OnlineKnapsackEV2021Sun} among others. In addition to the utilization rate of the knapsack, in \cite{OKPKeywordAuctions2008Zhou}, \cite{OnlineCloudPrices2017Zhangetal} upper and in \cite{OKPKeywordAuctions2008Zhou}, \cite{OnlineCloudPrices2017Zhangetal}, \cite{OnlineKnapsackEV2021Sun} lower limits on the value to weight ratio of the items, or upper limits on the item size, \cite{OnlineKnapsackDeparture2022Sunetal}, are assumed to be known.\\

Instead of using the knapsack utilization rates as the base for the thresholds, another approach for a threshold is to first observe some of the arrivals and then use the observed data to derive a threshold (see e.g. \cite{OnlineKnapsackDeparture2022Sunetal}, \cite{OGAPHistory2023Liuetal}, \cite{OSPhist2020Kaplanetal}, \cite{SecretaryProblem2007Babaioffetal}, \cite{SecretaryRobust2019Bradacetal}, \cite{OnlineSubmodularMaxConstraints2010Guptaetal}). This threshold is based on the values of the already observed items, similar to solution approaches for the classical secretary problem, in which the algorithm first observes a certain fraction of the online input, and the largest observed value is then used as the threshold for the remaining online input (see \cite{SecProbOpt1963Dynkin}). In \cite{OGAPHistory2023Liuetal} and \cite{OSPhist2020Kaplanetal}, the set of values, which should first be observed are referred to as the \textit{history set H}, and considered to be part of the input and not of the actual online problem, in which decisions are made. In case $H$ is small, the first items in the online phase are still only observed to gain additional information, similar to \cite{SecProbOpt1963Dynkin}. In \cite{OnlineSubmodularMaxConstraints2010Guptaetal}, various threshold-based online algorithms are proposed. Some directly rely on the original algorithm for the classical secretary problem, \cite{SecProbOpt1963Dynkin}, while others randomly select the number of items to first observe. In \cite{SecretaryProblem2007Babaioffetal} on the other hand, the set of items is split up into $L$ subsets, and after each round, the threshold is updated based on the solutions of the previous subset of items. A similar approach is used in \cite{SecretaryRobust2019Bradacetal}, in which some of the items arrive randomly within the time horizon $[0,1]$, and the split into subsets is done on the time horizon with updated thresholds for each interval.\\

A class of online problems, which closely resembles the considered online scheduling problem of a rolling horizon framework, are prophet inequality problems. In particular combinatorial prophet inequality problems, first introduced in \cite{CombiProphetInequ2017RubinsteinSingla} and improved in \cite{SubmodProphetInequality2021ChekuriLivanos}, are built upon a very similar structure. In these problems, the decision maker is faced with a sequence of $n$ items, drawn from $n$ known (discrete) distributions. At each iteration, an item is drawn from the corresponding distribution and presented to the decision maker, who must decide irrevocably whether to take the item or not, subject to matroid constraints. The goal is to maximize a submodular function $f$ over the set of chosen items. There are two main differences to the considered online scheduling problem of a rolling horizon. The first is the discrete nature of the underlying distributions. This difference may be overcome by approximating the continuous probability distributions by means of discrete distributions without losing too much information. The second difference lies within the knowledge of future distributions of the combinatorial prophet inequality problem. In our setting, we do not know the probability distributions, and therefore we are not able to directly apply the proposed algorithms to our case. In contrast to the previously considered literature, the algorithm is not based on an online threshold algorithm but is heavily dependent on \textit{Online Contention Resolution Schemes}, which is a rounding technique for online optimization problems, \cite{OCRS2016Feldmanetal}.\\

\subsection{Online Threshold Algorithms}\label{SS3OTA}
Summarizing the above-presented literature, the key similarity of most of the reviewed literature is the structure of how to decide whether to take item $t$ or not. This structure can be generalized and compressed to the following rather simple equation:
\begin{equation}\label{eq1Threshold}
	c_t \geq f(t,x)\tau,
\end{equation}
where $c_t$ is the value or contribution of item $t$, $f(t,x)$ is a factor, which may depend on time slot $t$ and further information $x$, such as how many items have already been taken or proposed, and $\tau$ is a threshold. Note that compared to the considered literature, the threshold is split into a threshold and a factor. If the value of the item exceeds the (factored) threshold, then the item is taken. Translating this simple decision process into the setting of an online scheduling problem for a rolling horizon, $\tau$ still may remain a threshold to be determined, while $f(t,x)$ may represent the possible dynamic nature of the threshold. On the other hand, the value $c_t$ of the current time slot $t$ may translate to the additional information, which the algorithm has access to in time slot $t$ compared to the last starting time slot. In contrast to the reviewed online literature, in our case, we are not simply given such a value $c_t$ but have to compute it.\\

Based on the principle of equation (\ref{eq1Threshold}) as well as the insights from the online literature, we identify three main lines of online algorithms, each with a distinct difference in either the value of a time slot or the threshold:
\begin{itemize}
	\item \textbf{Average-Realizations (AR) Approach}: Based on the longest path model introduced in Section \ref{S2Model} to determine optimal starting time slots, we define the individual contributions of each of the proposed optimal starting time slots. We use average realizations or predictions of the uncertainty as parameters for the path problem. After computing the individual contributions, we apply some function $g$ on the vector of contributions and use the corresponding result as the threshold. The contribution of the current time slot $t$ is computed based on the same type of data, but instead of using average realizations and forecasts, we can now use the actually observed realizations and current forecasts.
	\item \textbf{Historical-Realizations (HR) Approach}: Instead of using the longest path model with average realizations as the values, we may also use historical data to find optimal starting time slots for past days in hindsight. This historical data can be seen as samples from the underlying, unknown probability distribution, as used in prophet inequality problems, or as the set of items that should first be observed, as done in several online knapsack or secretary problems. Based on the chosen starting time slots, the individual contributions can easily be computed and used as the threshold. The contribution of time slot $t$ is again computed based on actual observations and updated forecasts.
	\item \textbf{Partial-Realizations (PR) Approach}: Instead of using the longest path model only once to compute individual contributions of optimal starting time slots, the model can also be used in an iterative online fashion. For each time slot, the model is run with updated real-time data, such as current forecasts and observed realizations of the previous time slots $s<t$. After solving the path problem with updated information, the decision of whether to start an iteration during the current time slot $t$ depends on the current solution. If node $t$ is part of this path, an iteration of the rolling horizon is started, else we repeat the process for the next time slot $t+1$.
\end{itemize}

In the following subsections, we focus on how to determine the contribution, threshold, and factor for time slot $t$ for each of the three main lines, given that the last starting time slot of the rolling horizon was a time slot $t_l<t$.

\subsubsection{Contribution}
In contrast to the simply observable value of an arriving item within an online knapsack, secretary, or prophet inequality problem, in our case, the value for the time slots needs to be computed. Hence, in the following, we refer to this value as the contribution of the current time slot. 
\begin{itemize}
	\item For the AR approach, we need to compute the contribution based on the differences in available information between the current time slot $t$ and the last starting time slot $t_l$. Depending on the application, the available information can either be in the form of updated forecasts or observed realizations of uncertainty since the last start.
	\item The contribution for the HR approach is the same as for the AR approach.
	\item The PR approach uses the optimal solution of the path model with updated values to derive the contribution. If the current time slot $t$ is part of the optimal path, we define the contribution as one, and if it is not, we simply set the contribution to zero.
\end{itemize}

\subsubsection{Threshold}\label{sss3Threshold}
All three main approaches (AR, HR, PR) derive their thresholds based on different data sets. While the AR approach is based on average values, the HR approach uses historical data to compute its threshold. The PR approach already uses its data within the computation of the contribution and can therefore make use of a very simple threshold. Within the following discussion of these thresholds, some (high-level) ideas from the reviewed literature can be identified. \\

\begin{itemize}
	\item The threshold for the AR approach is based on the optimal solution of the path model with average values as input parameters. Instead of directly using the nodes within the optimal path as fixed starting time slots of the iterations of the rolling horizon, as done in \cite{DynRHRMES2023Hoenenetal}, the AR approach instead uses the optimal objective value to compute its threshold. The objective value can be split up into individual contributions of the nodes within the optimal path. As the vector of individual contributions cannot directly be used as the online threshold, which consists of a single value, we need to apply some function on the vector to project it to a single number. Instead of using either the maximal, minimal, or average observation as the threshold, we test and analyze the impact of different percentiles as the threshold. 
	\item The threshold for the HR approach is also based on the individual contributions of the nodes of the optimal path of the path model. Instead of using average data for the values as done for the AR approach, the HR approach is based on historical data. We assume to have access to $N_{HR}$ such historical data sets, which represent 'similar' conditions as predicted for the current time horizon. This historical data is then used to compute optimal solutions in hindsight. The individual contributions of each of the $N_{HR}$ optimal paths are then grouped together and the percentile function is applied to generate one single value based on the vector of individual contributions.
	\item The threshold for the PR approach is relatively simple compared to the previous two approaches. As the contribution is either 0 or 1, represented by the optimal path, we simply set the threshold to 1. 
\end{itemize}

\subsubsection{Factor}
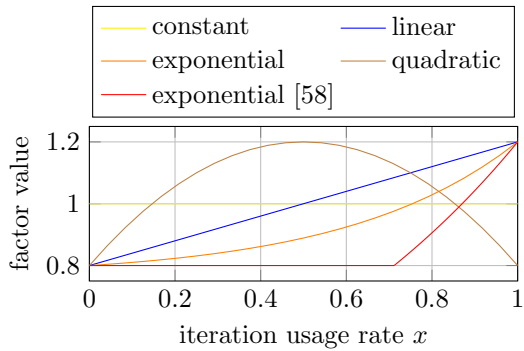
\begin{figure}
	\centering
	\pgfplotsset{every axis legend/.append style={at={(0.5, 1.78)}, anchor = north, legend columns = 2}}
	\begin{tikzpicture}
		\begin{axis}[xmin=0, xmax=1, ymin=0.75, ymax=1.25,
			xlabel={iteration usage rate $x$},
			ylabel={factor value},
			xtick distance = 0.2,
			ytick distance = 0.2,
			grid = both,
			major grid style = {lightgray},
			minor grid style = {lightgray!25},
			legend cell align={left},
			height=0.3\textwidth,
			width = 0.6\textwidth]
			\addplot[domain=0:1, yellow]{1};
			\addplot[domain=0:1, blue]{0.4*x+0.8};
			\addplot[domain=0:1, orange]{(0.4/(exp(2.5)-1))*exp(x*2.5)+((0.8*exp(2.5)-1.2)/(exp(2.5)-1))};
			\addplot[domain=0:1, brown]{-1.6*(x-0.5)^2+1.2};
			\addplot[domain=0:(1/(1+ln(1.2/0.8))), red]{0.8};
			\addplot[domain=(1/(1+ln(1.2/0.8))):1, red]{((1.2*exp(1)/0.8)^x)*(0.8/exp(1))};
			\legend{constant, linear, exponential, quadratic, exponential \cite{OKPKeywordAuctions2008Zhou}}
		\end{axis}
	\end{tikzpicture}
	\caption{Plot of the different factors, given a lower bound of $L=0.8$ and an upper bound of $U=1.2$}
	\label{figFactors}
\end{figure}

Based on the literature review of comparable online combinatorial optimization problems, most approaches use some form of dynamically changing threshold, often based on the utilization rate. The thresholds introduced in Section \ref{sss3Threshold} do not reflect this dynamic nature yet. Therefore, we have introduced the factor $f(t,x)$ in equation (\ref{eq1Threshold}) to account for that. In addition, insights into the structure of the underlying uncertainty as well as the shape of the contributions may be used to design 'tailor-made' factor functions $f(t,x)$. These factors $f(t,x)$ can be applied to the thresholds of the AR and HR approaches. Based on insights gained during the testing, the values of the factor functions should be centered around 1 to not change the threshold too much. To keep the functions general, we introduce a lower bound $L$ and an upper bound $U$ for the functions, whereby we assume $L<1<U$. The considered parameterized factor functions are then:
\begin{enumerate}
	\item \textbf{Constant}: A constant factor $f(t,x)=1$, leaving the 'pure' threshold, similar to ideas in secretary or prophet inequality problems (see e.g., \cite{ProphetSecretaryStaticThreshold2022ArnostiMa}, \cite{SecProbOpt1963Dynkin}).
	\item \textbf{Step-wise constant}: To discourage long periods without starting any iteration, we introduce a step-wise constant factor. This factor starts with a constant factor of 1 but decreases to a value of $L$ in case the time since the last starting time slot $t_l$ exceeds a certain setting-specific threshold $R$. This decrease should encourage to start another iteration, as the factored threshold also decreases, making it easier to start another iteration. Concrete, we use  
	\begin{equation}
		f(t,x) = 
		\begin{cases}
			1 & \text{if } t-t_l \leq R\\
			L & \text{if } t-t_l > R
		\end{cases}.
	\end{equation}
	\item \textbf{Linearly increasing}: Inspired by the linear threshold function presented in \cite{OMKPCompetetive2021Yangetal}, the following factor is considered: 
	\begin{equation}
		f(t,x) = (U-L) x + L.
	\end{equation}
	Note that this factor is not based on the time slot $t$, but rather on the iteration usage rate $x$, which represents the percentage of already used iterations during the time horizon. The main idea behind an increasing factor depending on the number of already started iterations is that in the beginning, it should be easier to start an iteration, but the more iterations have already started, the more difficult it should get and the 'pickier' the algorithm should be.
	\item \textbf{Exponentially increasing}: Much of the reviewed literature is based on an exponential threshold of some form, see \cite{OKPKeywordAuctions2008Zhou}, \cite{OnlineKnapsackDeparture2022Sunetal} among others. They are based on the utilization rate, a concept that we also use within this factor function. In contrast to exponential thresholds, which admit any item up to a given utilization rate (see e.g., \cite{OKPKeywordAuctions2008Zhou}) we slightly lower the threshold in the beginning and then directly increase it. This leads to the following factor function 
	\begin{equation}
		f(x) = a \exp(c x) + b,
	\end{equation}
	where $a, b$ and $c$ are chosen such that $f(0)=L$ and $f(1)=U$. In Figure \ref{figFactors}, $c$ is chosen to be 2.5, while $L=0.8$ and $U=1.2$.
	\item \textbf{Quadratically increasing and decreasing}: The idea of starting with a lower threshold and then making it more difficult to start new iterations, as done for the linear and the exponential factors, has been inspired by the reviewed theoretical work on online-competitive algorithms. In practice, on the other hand, this increasing factor leads to situations, in which not all of the possible iterations are used, which leaves potential unused. To counter this, we let the quadratic factor not continue to increase with the utilization rate $x$, but decrease it again for $x>0.5$. We define the function as follows:
	\begin{equation}
		f(x) = a (x-0.5)^2 + b,
	\end{equation}
	where $a$ and $b$ are again chosen such that $f(0)=f(1)=L$ and $f(0.5)=U$.
	\item \textbf{Factored threshold based on \cite{OKPKeywordAuctions2008Zhou}}: Next to concepts from literature, we may also directly use a (slightly modified) version of the threshold function presented in \cite{OKPKeywordAuctions2008Zhou}. However, as the assumption of a strictly positive upper and lower limit on the value does not hold in this study, we simply use the minimal and maximal contribution values of the HR knapsack problem as the lower and upper limit.
\end{enumerate}

\section{Application to an Energy Management Problem Under Uncertainty}\label{S4ApplicationScenario}

Within this section, we use an energy management problem under uncertainty for a microgrid as an example of applying the introduced online algorithms for the scheduling of a rolling horizon. We shortly introduce the energy planning problem, including the involved participants and components, and then focus on the parameters needed as input for the models introduced in Section \ref{S2Model}. Within the energy management problem, we apply robust optimization to deal with uncertainty in the data but treat the robust formulation as a black-box within this work. As this work is a continuation of earlier work, we refer the interested reader to \cite{DynRHRMES2023Hoenenetal}, Sections 2 and 3, for further details regarding the setting of the energy management problem. In addition, a comprehensive explanation of the robust energy management formulation and the classical and dynamic, yet offline rolling horizon frameworks can be found there. We conclude the section with a detailed overview of the needed input parameter for the online algorithms.

\subsection{Robust Energy Management Problem}
\subsubsection{Setting}
The considered dynamic problem is an energy management problem for a microgrid under uncertainty, in which households may have controllable devices, such as EVs, batteries, or PV systems. Hereby, the time horizon is discretized and spans multiple days. Furthermore, two different electricity markets are available to purchase or sell electricity. We assume that a microgrid operator (MGO) acts as an aggregator for the microgrid and is therefore responsible for the buying and selling decisions at the markets.\\

The two electricity markets are the \textit{day-ahead market} and the \textit{intraday market}, which work as their names indicate. The decisions for the day-ahead market have to be made and submitted the day before the actual usage. Given the time distance between submitting the buying and selling decisions and their actual implementation, deviations from the planned energy usage may occur. The MGO can use the intraday market to react to these deviations by deciding just before a time slot $t$ how much to buy or sell at the intraday market (additionally to the day-ahead market). The buying and selling prices for time slot $t$ in this market are denoted with $\pi^{ID,buy}_t$ and $\pi^{ID,sell}_t$ and with $\pi^{DA}_t$ for the day-ahead market.\\

The considered devices and loads of the household are the following:
\begin{itemize}
	\item Prosumer load: Each household has an inflexible load profile, which has to be served. We do not assume that any of this load can be curtailed or shifted to other time slots.  
	\item PV system: Households may be equipped with a PV system. Let $\hat{p}^{PV,j}_t$ denote the produced electricity of PV system $j$ during time slot $t$. The MGO can curtail the generation if needed. 
	\item Battery: A battery can be used to shift the demand or production of electricity through time.
	\item EV: The EV works in a similar way as the battery in that it can be charged and discharged. A major difference is the availability of the EV, which is limited, as well as the EV demand, which reduces its state of charge.
\end{itemize}

\subsubsection{Uncertainty}
For the above-presented setting of a microgrid energy management problem, we consider multiple sources of uncertainty. We first reason why parameters are assumed to be uncertain, and then provide the corresponding notation and uncertainty sets:
\begin{itemize}
	\item Prosumer load: The household load of an individual household is heavily influenced by often spontaneous decisions and actions of its residents. Hence, it is not possible to perfectly predict individual household loads for even short-term time horizons.
	\item PV generation: Although PV forecasting has continued to improve over the last years, still deviations between the predicted value and the actual generations do appear. A general trend is that short-term forecasting often yields better results than long-term forecasting of PV generation \cite{PVforecasting2019Hanetal}. We include this insight into the modeling of the PV uncertainty in the following way. We first split up the uncertain parameter $\hat{p}^{PV}_t$ into a known and an uncertain part,
	\begin{equation}
		\hat{p}^{PV}_t = p^{PV}_t(1+\alpha^{PV}_tu^{PV}_t),
	\end{equation}
	where $p^{PV}_t$ corresponds to the predicted value, while $\alpha^{PV}_tu^{PV}_t$ represents the uncertain part. This uncertain part can be further split up into $u^{PV}_t$, which is a random variable, modeling the true realization of the uncertainty, and $\alpha^{PV}_t$, which defines how large the uncertainty can potentially be. Note that $\alpha^{PV}_t$ depends on time slot $t$, which is how the different accuracies of the forecast are included in the model. In particular during the rolling horizon framework, $\alpha^{PV}_t$ and $p^{PV}_t$ for a given future time slot $t$ can change depending on when the forecast is made. For using robust optimization, we define the uncertainty set for $u^{PV}_t$ as a budget uncertainty set:
	\begin{equation}
		U^{PV} = \left\lbrace u \in \mathbb{R}^{\vert \mathcal{T} \vert} \vert \Vert u \Vert_{\infty} \leq 1, \Vert u \Vert_1 \leq \Gamma^{PV} \right\rbrace.
	\end{equation}
	In addition, we assume that the PV forecasts are time-dependent, that is they improve over time, meaning that the uncertainty sets get smaller, the closer we are to the corresponding time slot. For this, we need to introduce another index for the PV data to indicate when the forecast has been made. Let $p^{PV}_{t,s}$ denote the PV generation forecast for time slot $t$, made at time slot $s$. The same applies to $\alpha^{PV}_{t,s}$, which is predicted at time slot $s$ for time slot $t$. In case only one time slot index is present, we assume w.l.o.g. that the initial forecast of the beginning of the time horizon is used. One important assumption on the resulting sequence of PV forecasts for a fixed time slot $t$ is that they are contained in each other, i.e.,
	\begin{equation}\label{PVlowerChain}
		p^{PV}_{t,s}(1-\alpha^{PV}_{t,s}) \geq p^{PV}_{t,l}(1-\alpha^{PV}_{t,l}),
	\end{equation}
	and
	\begin{equation}\label{PVupperChain}
		p^{PV}_{t,s}(1+\alpha^{PV}_{t,s}) \leq p^{PV}_{t,l}(1+\alpha^{PV}_{t,l}),
	\end{equation}
	for time slots $l \leq s \leq t$. This ensures that the uncertainty intervals for a specific time slot $t$ can only improve.
	\item EV demand: Although the EV demand for a known trip can be estimated quite well, in practice various unknown factors, such as vehicle heating or cooling, traffic jams or small detours can affect the actual demand. Hence, we also consider the EV demand to be uncertain.
	\item EV arrival and departure times: Even though most people have some level of routine in their daily working patterns, unforeseen events, such as traffic jams, or an unplanned detour to the supermarket can always change arrival and departure times.
	\item Market prices: As the prices of the electricity markets are based on the actual demand and supply for a time slot, we assume that the exact prices are not known in advance. 
\end{itemize}

\subsubsection{Robust Energy Management Approach}
For the introduced setting, the goal of the MGO is to minimize the electricity costs of the microgrid, while maintaining a feasible schedule at all times. For this, the MGO needs to decide when and how much electricity to buy and sell at the two considered electricity markets. In addition, it also needs to decide on the charging and discharging schedules of batteries and EVs, while satisfying the regular household demand. All these decisions have to be made under the assumption that the exact values of demand, generation, and prices are unknown beforehand.

\subsection{Application to Online Framework}\label{Ss4.2Application}
Determining the values and contributions needed for the online scheduling framework for this robust energy management problem is mainly based on the improvements in PV forecast over time. The conservative approach of robust optimization enforces that the obtained solution is feasible w.r.t. all possible realizations of the uncertain parameter. In the case of the uncertain PV generation, this translates into using at most the lower limit of the PV uncertainty interval. With each iteration, we get access to updated forecasts, and thereby also to (improved) lower limits of the PV uncertainty intervals. For the three considered different online scheduling algorithms, we have to estimate how much this lower limit will improve. This results in the following values and parameters:
\begin{itemize}
	\item AR approach: Assuming average contributions, we use historical knowledge on how the uncertainty sets for a given time slot $t$ shrink over time, as assumed by equations (\ref{PVlowerChain}) and (\ref{PVupperChain}). Let $\beta^{PV}_{l,t} \in \left[ 0,1\right]$ be the reduction factor of the PV uncertainty interval for time slot $l$, expressing an improved forecast at time $t$ compared to the initial (long-term) forecast at time $0$. Then the average improvement of the lower limit of the PV uncertainty interval can be expressed as $p^{PV}_{l} \alpha^{PV}_{l} \beta^{PV}_{l,t}$. Hence, the average improvement for starting an iteration at time slot $t$ in relation to the last start at time slot $s$ is given by
	\begin{equation}
		c_{s,t} = \sum_{l=t}^{\vert \mathcal{T} \vert} p^{PV}_{l}\alpha^{PV}_{l} \big( \beta^{PV}_{l,t} - \beta^{PV}_{l,s} \big).
	\end{equation}
	This improvement can then be seen as the additional available PV generation compared to time slot $s$. These values are used as the directed arc lengths for the longest path model.\\
	
	Within the online process of deciding whether to start an iteration at time slot $t$ or not, we need to compare the threshold, which is (loosely) based on the above-presented values, with the contributions of time slot $t$, assuming we would start an iteration. Therefore, we use the current data and forecasts of time slot $t$ and compute the improvements over the forecasts made at the last starting time slot $t_l$, resulting in 
	\begin{equation}\label{Contribution}
		\Tilde{c}_t = \sum_{s=t}^{T}(p^{PV}_{t_l,s}(1-\alpha^{PV}_{t_l,s}) - p^{PV}_{t,s}(1-\alpha^{PV}_{t,s})),
	\end{equation}
	as the contribution of the current time slot $t$.
	
	\item HR approach: We translate each of the given data samples into a longest path problem and thereby determine the optimal starting time slots of the iterations in hindsight. Given that, for each data sample, we know all historical PV forecasts, we compute the value $c_{s,t}$ for time slots $s<t \in \mathcal{T}$ by
	\begin{equation}
		c_{s,t} = \sum_{l=t}^{\vert \mathcal{T} \vert} \big( p^{PV}_{l,t}(1-\alpha^{PV}_{l,t}) - p^{PV}_{l,s}(1-\alpha^{PV}_{l,s}) \big).
	\end{equation}
	For a fixed time slot $l>t$, this can be seen as the improvement in the lower bound of the uncertainty interval of the PV production given the new starting time slot $t$ over the previous prediction made at time slot $s$.\\
	
	Remind that for the HR approach, the same threshold (\ref{Contribution}) as for the AR approach is used. 
	
	\item PR approach: For the current time slot $t$, we combine the data of the average realizations with observations of the past. Let $t_l$ again be the last realized starting time slot. Then, we have four different cases:
	\begin{enumerate}
		\item For arc $(t_l,t)$, we simply use the predicted forecasts, and subtract the lower limits of the predicted PV uncertainty intervals: 
		\begin{equation}
			c_{t_l,t} = \sum_{k=t}^{\vert \mathcal{T} \vert} \big( p^{PV}_{k,t} (1 - \alpha^{PV}_{k,t}) - p^{PV}_{k,t_l} (1 - \alpha^{PV}_{k,t_l}) \big).
		\end{equation}
		Due to constraint (\ref{PVlowerChain}), the length of the arc $(t_l,t)$ is always positive. 
		\item For arcs $(t_l, l)$ with $l>t$, we again apply the idea of average reductions for future forecasts, while for time slot $t_l$, we can simply use the lower limits of the predicted PV uncertainty intervals. Hence, we have 
		\begin{equation}
			c_{t_l,l} = \sum_{k=l}^{\vert \mathcal{T} \vert} \big( p^{PV}_{k,t} (1 - \alpha^{PV}_{k,t} + \alpha^{PV}_{k,t} (\beta^{PV}_{k,l}-\beta^{PV}_{k,t})) - p^{PV}_{k,t_l} (1 - \alpha^{PV}_{k,t_l}) \big),
		\end{equation}
		where the $p^{PV}_{k,t} (1 - \alpha^{PV}_{k,t} + \alpha^{PV}_{k,t} (\beta^{PV}_{k,l}-\beta^{PV}_{k,t}))$ corresponds to the estimated lower limit of the PV uncertainty interval for time slot $k$, made at time slot $l$. Note that instead of using the long-term forecast made at time 0 as the basis for the improvement factors $\beta$, we use the most recent forecast from time slot $t$, which may already improve upon the forecast made at time slot $0$. To account for that, we subtract the improvement factor $\beta^{PV}_{k,t}$ at time $t$, resulting in an updated improvement factor of $(\beta^{PV}_{k,l}-\beta^{PV}_{k,t})$. Hence, the lower limit of the PV uncertainty interval is based on the most recent lower limit estimation from time slot $t$, but adjusted for the average reduction factor. Using equation (\ref{PVlowerChain}) and $0 \leq \beta_{k,t} < \beta_{k,l} \leq 1$, we have $c_{t_l,l}\geq 0$ for $l>t$.
		\item For arcs $(t,l)$ with $t<l$, we use the current prediction of time slot $t$ and assume average realizations for future time slots
		\begin{equation}
			\begin{split}
				c_{t,l} = & \sum_{k=l}^{\vert \mathcal{T} \vert} \big( p^{PV}_{k,t} (1 - \alpha^{PV}_{k,t} + \alpha^{PV}_{k,t} (\beta^{PV}_{k,l}-\beta^{PV}_{k,t})) - p^{PV}_{k,t} (1 - \alpha^{PV}_{k,t}) \big) \\
				= & \sum_{k=l}^{\vert \mathcal{T} \vert} \big( p^{PV}_{k,t} \alpha^{PV}_{k,t} (\beta^{PV}_{k,l}-\beta^{PV}_{k,t}) \big).
			\end{split}
		\end{equation}
		\item For the remaining arcs $(s,l)$, with $t < s < l$, we again assume the average realizations of the AR approach. That is, we use the current forecast at time slot $t$ for future time slots $k$ to compute how large the improvement of starting an iteration at time slot $l$ is over time slot $s$. Hence, for time slots $t\leq s<l$ we have
		\begin{equation}
			c_{s,l} = \sum_{k=l}^{\vert \mathcal{T} \vert} p^{PV}_{k,t}\alpha^{PV}_{k,t} \big( \beta^{PV}_{k,l} - \beta^{PV}_{k,s} \big).
		\end{equation}
	\end{enumerate}
	
\end{itemize}

For all three approaches, let $\mathcal{M}$ be defined as the set of time slots, in which the decisions regarding the day-ahead market need to be submitted. Within this setting, these time slots are 12 pm every day. Note that also the observations of the uncertain EV demand could be used within the values and thresholds of the online algorithms, as done in \cite{DynRHRMES2023Hoenenetal}. We refrain from including the EV demand in the threshold valuation design as the realizations of EV demand are not time-critical.

\section{Numerical Results and Analysis}\label{S5Analysis}

In this section, we test, analyze and discuss the various online scheduling approaches, presented in Section \ref{S3OnlineAlg}. We first introduce the simulation setting and then proceed with the analysis of the online threshold algorithm. We deliberately focus only on the online algorithm without the underlying robust energy management approach, as introduced in Section \ref{S3OnlineAlg} to gain insights into the working of the proposed algorithms. We start with a detailed analysis of the impact of the threshold on the solution of the basic online threshold algorithm and then analyze the impact of the various parameter choices. Based on these results, we propose and further test only selected parameter combinations for the case of including the robust energy management approach.

\subsection{Simulation}
The simulations are all based on a microgrid of 20 households, equipped with 17 PV systems, 15 EVs and one jointly used battery system. The considered time horizon spans 3 days and consists of 288 time slots of 15 minutes. 

\subsubsection{Data}
To model the household load, we use average Dutch load profiles from the 12th to the 14th of April 2021 \cite{NEDUDemandProfiles}. These profiles are scaled to a household with a yearly demand of 3,500 kWh, which results in a daily consumption of 8 to 10 kWh per household. All PV systems are modeled equal, with a production of up to 11 kWh per day. The generation profile is based on a sunny day. The EVs are based on the VW ID.3, with a battery capacity of 58 kWh and charging and discharging limits of 11 kW. The charging and discharging efficiencies are assumed to be 95\%. The EV demand is computed based on daily trips between 20 and 70 km and an electricity usage of 18 kWh / 100 km. The jointly used battery system consists of 3 interconnected Tesla Powerwall modules with an aggregated battery capacity of 42 kWh \cite{TeslaPowerWall}. The charging and discharging power limits are 15 kW, and the efficiencies are again set to 95\%. W.l.o.g., we assume that the initial SoC of EVs and the battery is 0. The day-ahead electricity prices for the considered time horizon are taken from \cite{DAMarketPrices}, while the intraday market prices are taken from the Dutch TSO TenneT \cite{IDMarketPrices}.

\subsubsection{Uncertainty Sets}
The used parameters for the considered uncertainty, as introduced in Section \ref{S4ApplicationScenario}, are: 
\begin{itemize}
	\item The household load uncertainty is defined as $\alpha^L=0.2$.
	\item The EV demand uncertainty is $\alpha^{EV}=0.1$.
	\item The uncertainty of the electricity prices is $\alpha^{DA}=0.15$ for the day-ahead market and $\alpha^{ID}=0.35$ for the intraday market.
	\item The long-term forecast uncertainty for the PV generation is equal to $\alpha^{PV}=0.25$, but there is a reduction in uncertainty for short-term forecasts. Within this work, we assume that the uncertainty interval predicted at time slot $t-s$ for the generation at time slot $t$ is given by
	\begin{equation}\label{PVUncertaintyInterval}
		\left[ p^{PV}_{t,t-s} (1-\alpha^{PV}_{t,t-s}), p^{PV}_{t,t-s} (1+\alpha^{PV}_{t,t-s}) \right].
	\end{equation}
	Hereby, $\alpha^{PV}_{t,t-s}$ reduces by a fixed percentage the smaller $s$ gets, as already described in Section \ref{S4ApplicationScenario}. This reduction ranges from 0\% for $s>8$, over 1\% for $s=8$ to nearly 70\% for $s=0$. This implies that the uncertainty of the forecast for the coming time slot has decreased by around 70\% compared to the initial long-term forecast.
\end{itemize}

\subsubsection{Uncertainty Realizations}
Next to the time dependency of some uncertainty sets, another important factor in the evaluation of the algorithms is the underlying probability distribution of the uncertainty realizations. In most literature on robust optimization, a uniform distribution is used to draw the actual realizations (see e.g., \cite{AROUnitCommitment2013Bertsimasetal}, \cite{ROV2G2015BaiQiao}). Such a uniform distribution seems to be matching best with the assumption of having no further knowledge on the underlying distribution apart from the support. Unless mentioned otherwise, the realizations within this work are all drawn from a uniform distribution on the interval $\left[ -1,1 \right]$.\\

While the usage of a uniform distribution is common in the context of robust optimization, in practice, error realizations do not always follow a uniform distribution. A very popular choice is the normal distribution (see e.g., \cite{RHdynamicEM2022Yodasetal}, \cite{PVerrorDist2020Tanetal}, \cite{PVerrorDist2014Ziadietal}). We use a truncated normal distribution on the interval $\left[ -1,1 \right]$ with mean 0 and standard deviation 0.4 to model the realizations for the PV uncertainty. We stick to the uniform distribution for the remaining uncertain parameters to better identify, how well the online scheduling approach works for different uncertainty distributions.\\

To also be able to analyze the performance of the online (and dynamic) algorithms under a non-zero mean error distribution, we also consider a truncated normal distribution with a shifted mean for the PV generation. Note, that the term non-zero mean does not imply that the forecasts of PV are off, but that the uncertainty is asymmetrically distributed around the predicted value. This may reflect that forecasting algorithms are able to predict the PV generation quite well if there are no clouds, but if there are a few clouds, which may block direct irradiation for short time periods, this will lead to larger (negative) deviations from the predicted value. Concisely, the uncertain PV generation is assumed to be given by a shifted normal distribution with a mean of 0.6 and a standard deviation of 0.4.

\subsection{Analysis Online Threshold Algorithms}\label{S52AOTA}
In this section, we focus on the analysis of the online threshold algorithms without the underlying robust energy management approach. This implies that we only use the information gains, as defined in Section \ref{Ss4.2Application}. We first test the impact of the threshold on the outcome of the online threshold algorithms, before analyzing the impact of the various individual parameter choices, such as the percentiles, the factor, or the different approaches presented in Section \ref{S3OnlineAlg}.\\ 

\subsubsection{Threshold}
We focus on the impact of the threshold on the outcome of a basic online threshold algorithm with a constant threshold. We do not yet analyze the impact of the various parameters on the results of the algorithms but rather show how the online threshold algorithm works and how we can evaluate and compare its performance.\\

\begin{figure}
	\centering
	\pgfplotsset{every axis legend/.append style={at={(0.5, 1.32)}, anchor = north, legend columns = 2}}
	\begin{tikzpicture}
		\begin{axis}[xmin=0, xmax=97, ymin=0, ymax=2.9,
			xlabel={Time slot of the day},
			ylabel={add. PV $[kWh]$},
			grid = both,
			minor tick num = 1,
			major grid style = {lightgray},
			minor grid style = {lightgray!25},
			legend cell align={left},
			height=0.45\textwidth,
			width = 0.7\textwidth]
			\addplot[red]table[col sep = comma, x={i}, y={norm}]{SimpleThreshold.csv};
			\addplot[black]table[col sep = comma, x={i}, y={once}]{SimpleThreshold.csv};
			\addplot[green, domain=0:97]{1.4668};
			\addplot[blue, samples at={35}, only marks, mark = x, mark size = 2]{1.484};
			\addplot[blue, samples at={39}, only marks, mark = x, mark size = 2]{1.767};
			\addplot[blue, samples at={42}, only marks, mark = x, mark size = 2]{1.906};
			\addplot[blue, samples at={45}, only marks, mark = x, mark size = 2]{2.098};
			\addplot[blue, samples at={47}, only marks, mark = x, mark size = 2]{1.737};
			\addplot[green, samples at={48}, only marks, mark = x, mark size = 2]{0.832};
			\addplot[blue, samples at={50}, only marks, mark = x, mark size = 2]{1.917};
			\addplot[blue, samples at={52}, only marks, mark = x, mark size = 2]{1.663};
			\addplot[blue, samples at={55}, only marks, mark = x, mark size = 2]{1.895};
			\addplot[blue, samples at={58}, only marks, mark = x, mark size = 2]{1.934};
			\addplot[blue, samples at={60}, only marks, mark = x, mark size = 2]{1.556};
			\addplot[blue, samples at={64}, only marks, mark = x, mark size = 2]{1.543};
			\legend{Contributions, Fully Static, Threshold, Starting Points}
		\end{axis}
	\end{tikzpicture}
	\caption{Contributions per time slot for online approach (constant factor, including starting times slots) and fully static approach (solves the model only once). The iteration at time slot 48 is part of the set $\mathcal{M}$ and therefore ignores the threshold.}
	\label{FigNormContri}
\end{figure}
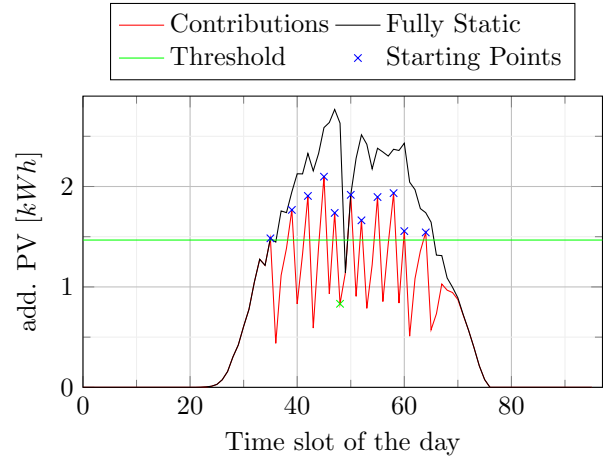

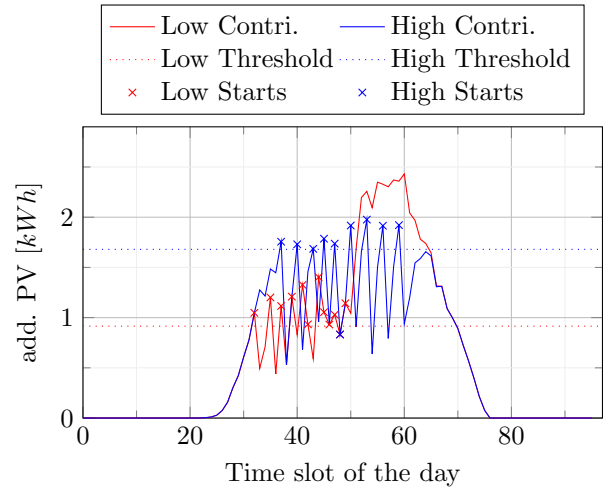
\begin{figure}
	\centering
	\pgfplotsset{every axis legend/.append style={at={(0.5, 1.42)}, anchor = north, legend columns = 2}}
	\begin{tikzpicture}
		\begin{axis}[xmin=0, xmax=97, ymin=0, ymax=2.9,
			xlabel={Time slot of the day},
			ylabel={add. PV $[kWh]$},
			grid = both,
			minor tick num = 1,
			major grid style = {lightgray},
			minor grid style = {lightgray!25},
			legend cell align={left},
			height=0.45\textwidth,
			width = 0.7\textwidth]
			\addplot[red]table[col sep = comma, x={i}, y={low}]{SimpleThreshold.csv};
			\addplot[blue]table[col sep = comma, x={i}, y={high}]{SimpleThreshold.csv};
			\addplot[red, dotted, domain=0:97]{0.915};
			\addplot[blue, dotted, domain=0:97]{1.68};
			\addplot[red, samples at={32}, only marks, mark = x, mark size = 2]{1.048};
			\addplot[blue, samples at={37}, only marks, mark = x, mark size = 2]{1.756};
			\addplot[red, samples at={35}, only marks, mark = x, mark size = 2]{1.201};
			\addplot[red, samples at={37}, only marks, mark = x, mark size = 2]{1.116};
			\addplot[red, samples at={39}, only marks, mark = x, mark size = 2]{1.209};
			\addplot[red, samples at={41}, only marks, mark = x, mark size = 2]{1.326};
			\addplot[red, samples at={42}, only marks, mark = x, mark size = 2]{0.933};
			\addplot[red, samples at={44}, only marks, mark = x, mark size = 2]{1.403};
			\addplot[red, samples at={45}, only marks, mark = x, mark size = 2]{1.054};
			\addplot[red, samples at={46}, only marks, mark = x, mark size = 2]{0.931};
			\addplot[red, samples at={47}, only marks, mark = x, mark size = 2]{1.029};
			\addplot[red, samples at={48}, only marks, mark = x, mark size = 2]{0.832};
			\addplot[red, samples at={49}, only marks, mark = x, mark size = 2]{1.143};
			\addplot[blue, samples at={40}, only marks, mark = x, mark size = 2]{1.730};
			\addplot[blue, samples at={43}, only marks, mark = x, mark size = 2]{1.685};
			\addplot[blue, samples at={45}, only marks, mark = x, mark size = 2]{1.786};
			\addplot[blue, samples at={47}, only marks, mark = x, mark size = 2]{1.737};
			\addplot[blue, samples at={48}, only marks, mark = x, mark size = 2]{0.832};
			\addplot[blue, samples at={50}, only marks, mark = x, mark size = 2]{1.917};
			\addplot[blue, samples at={53}, only marks, mark = x, mark size = 2]{1.976};
			\addplot[blue, samples at={56}, only marks, mark = x, mark size = 2]{1.915};
			\addplot[blue, samples at={59}, only marks, mark = x, mark size = 2]{1.921};
			\legend{Low Contri., High Contri., Low Threshold, High Threshold, Low Starts, High Starts}
		\end{axis}
	\end{tikzpicture}
	\caption{Contributions per time slot for online approach with low and high constant threshold to demonstrate the impact of the threshold on the solution}
	\label{FigExtremeContri}
\end{figure}

We start with a visual representation of the achieved results of the constant threshold-based online algorithm. Figure \ref{FigNormContri} displays the development of the contribution per time slot (see Equation (\ref{Contribution})), for one day based on a constant threshold. The individual contribution represents the additional PV generation, which can be used due to improved PV forecasts compared to the start of the last iteration. The figure also shows the used threshold as well as the chosen starting time slots. As can be seen, the algorithm starts an iteration whenever it exceeds the threshold and therefore works as intended. However, this may change if the threshold takes more extreme values. Figure \ref{FigExtremeContri} shows that if the threshold is too low, all iterations are used but are distributed very early during the day, whereas for a higher threshold, the starting time slots are distributed more evenly during the day, but do not use all possible iterations. As iterations starting very early or late during the day are likely to achieve smaller contributions, the choice of the threshold seems to be of great importance to achieve a well-balanced solution, in which (nearly) all iterations are used, but are also well spread among the time horizon.\\ 

Next to the visual analysis of the starting time slots, we can calculate the objective value of the online algorithms, given by the sum of collected contributions. The resulting objective values for the three scenarios in Figures \ref{FigNormContri} and \ref{FigExtremeContri}, confirm the initial intuition. While the threshold displayed in Figure \ref{FigNormContri} has an objective value of 20.333 kWh, the thresholds from Figure \ref{FigExtremeContri} only achieve an objective of 13.218 kWh (low), respectively 17.256 kWh (high). Hence, even a small change in the threshold can lead to a large change in the objective value.\\

\begin{figure}
	\centering
	\pgfplotsset{every axis legend/.append style={at={(0.5, 1.22)}, anchor = north, legend columns = 3}}
	\begin{tikzpicture}
		\begin{axis}[xmin=0, xmax=97, ymin=0, ymax=2.9,
			xlabel={Time slot of the day},
			ylabel={add. PV $[kWh]$},
			grid = both,
			minor tick num = 1,
			major grid style = {lightgray},
			minor grid style = {lightgray!25},
			legend cell align={left},
			height=0.45\textwidth,
			width = 0.7\textwidth]
			\addplot[red]table[col sep = comma, x={i}, y={opt}]{SimpleThreshold.csv};
			\addplot[blue, samples at={35}, only marks, mark = x, mark size = 2]{1.484};
			\addplot[green, samples at={32}, only marks, mark = x, mark size = 2]{1.048};
			\addplot[blue, samples at={39}, only marks, mark = x, mark size = 2]{1.767};
			\addplot[blue, samples at={42}, only marks, mark = x, mark size = 2]{1.906};
			\addplot[blue, samples at={45}, only marks, mark = x, mark size = 2]{2.098};
			\addplot[blue, samples at={48}, only marks, mark = x, mark size = 2]{2.099};
			\addplot[blue, samples at={50}, only marks, mark = x, mark size = 2]{1.917};
			\addplot[blue, samples at={52}, only marks, mark = x, mark size = 2]{1.663};
			\addplot[blue, samples at={55}, only marks, mark = x, mark size = 2]{1.895};
			\addplot[blue, samples at={58}, only marks, mark = x, mark size = 2]{1.934};
			\addplot[blue, samples at={60}, only marks, mark = x, mark size = 2]{1.556};
			\addplot[blue, samples at={64}, only marks, mark = x, mark size = 2]{1.543};
			\addplot[blue, samples at={67}, only marks, mark = x, mark size = 2]{1.029};
			\addplot[green, samples at={33}, only marks, mark = x, mark size = 2]{1.275};
			\addplot[green, samples at={34}, only marks, mark = x, mark size = 2]{1.215};
			\addplot[green, samples at={37}, only marks, mark = x, mark size = 2]{1.116};
			\addplot[green, samples at={38}, only marks, mark = x, mark size = 2]{1.385};
			\addplot[green, samples at={41}, only marks, mark = x, mark size = 2]{1.326};
			\addplot[green, samples at={44}, only marks, mark = x, mark size = 2]{1.403};
			\addplot[green, samples at={47}, only marks, mark = x, mark size = 2]{1.737};
			\addplot[green, samples at={49}, only marks, mark = x, mark size = 2]{1.143};
			\addplot[green, samples at={54}, only marks, mark = x, mark size = 2]{1.221};
			\addplot[green, samples at={57}, only marks, mark = x, mark size = 2]{1.428};
			\addplot[green, samples at={62}, only marks, mark = x, mark size = 2]{1.084};
			\addplot[green, samples at={63}, only marks, mark = x, mark size = 2]{1.358};
			\legend{Optimal, OptStarts, ConstStarts}
		\end{axis}
	\end{tikzpicture}
	\caption{Contributions of an optimal solution, with OptStarts denoting the optimal starting time slots of the rolling horizon, while ConstStarts denotes the time slots, in which the contribution is larger than the smallest optimal contribution, showing that a constant threshold policy does not lead to an optimal solution}
	\label{FigContriOptConstant}
\end{figure}
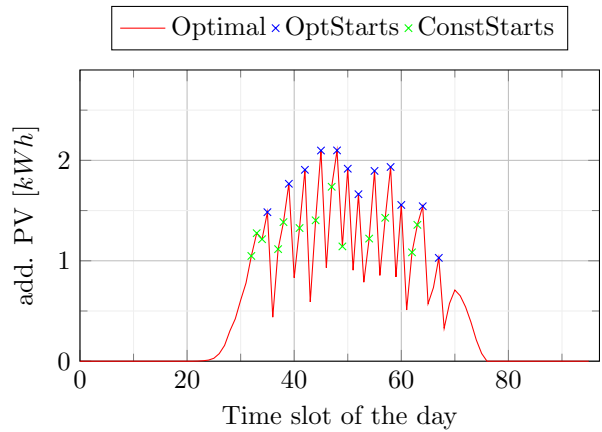

The above comparison gives some insights into which threshold performs better, but it does not indicate how well the best threshold might perform. For this, we may use an upper bound on the objective value. Within the area of online (combinatorial) optimization, often the optimal objective value of the offline scenario, in which all information is known beforehand, is used as a bound. In our case, we use the path model with the actual forecasts to compute the optimal starting time slots of an instance in hindsight. Figure \ref{FigContriOptConstant} shows the optimal distribution of starting time slots and the corresponding individual contributions given the realizations of uncertainty for the considered instance. The optimal offline objective of the instance is 20.891 kWh, showing an 'optimality gap' of less than 2.7\% for the threshold used in Figure \ref{FigNormContri}.\\

Figure \ref{FigContriOptConstant} also shows that we cannot hope for a constant threshold, which leads to an optimal solution for all instances. Taking the smallest individual contribution of the optimal solution as the constant threshold would result in a solution similar to that of the low threshold presented in Figure \ref{FigExtremeContri}. Generalizing this insight implies that there does not exist an optimal convex threshold policy for the online $k$-edge longest path problem. \\

\subsubsection{Factor}
Although it is not possible to derive an optimal online threshold policy to our problem, in practice it has been shown that simple online threshold algorithms often achieve good results (see e.g., \cite{ProphetSecretaryStaticThreshold2022ArnostiMa}, \cite{OMKPCompetetive2021Yangetal}). In the following, we, therefore, test the impact of the factor choice on the outcome of the online threshold algorithms.\\

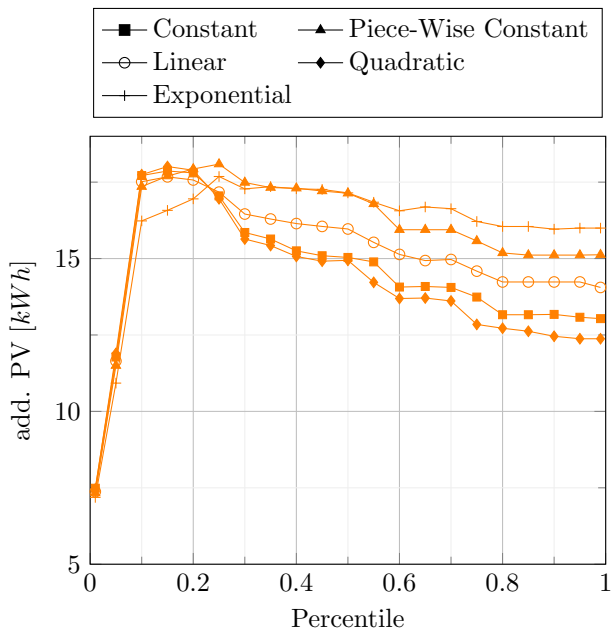
\begin{figure}
	\centering
	\pgfplotsset{every axis legend/.append style={at={(0.5, 1.3)}, anchor = north, legend columns = 2}}
	\begin{tikzpicture}
		\begin{axis}[xmin=0, xmax=1, ymin=5, ymax=19,
			xlabel={Percentile},
			ylabel={add. PV $[kWh]$},
			grid = both,
			minor tick num = 1,
			major grid style = {lightgray},
			minor grid style = {lightgray!25},
			legend cell align={left}]
			\addplot[black, mark = square*]{-10};
			\addplot[black, mark = triangle*]{-10};
			\addplot[black, mark = o]{-10};
			\addplot[black, mark = diamond*]{-10};
			\addplot[black, mark = +]{-10};
			\addplot[orange, mark = square*, mark size = 1.5]table[col sep = comma, x={P}, y={800}]{FinalContriUniform.csv};
			\addplot[orange, mark = triangle*]table[col sep = comma, x={P}, y={804}]{FinalContriUniform.csv};
			\addplot[orange, mark = o]table[col sep = comma, x={P}, y={801}]{FinalContriUniform.csv};
			\addplot[orange, mark = diamond*]table[col sep = comma, x={P}, y={805}]{FinalContriUniform.csv};
			\addplot[orange, mark = +]table[col sep = comma, x={P}, y={806}]{FinalContriUniform.csv};
			\legend{Constant, Piece-Wise Constant, Linear, Quadratic, Exponential}
		\end{axis}
	\end{tikzpicture}
	\caption{Contributions of uniform AR online threshold algorithms for various factors (marks) and percentiles (x-axis) for a time horizon of three days with $k=12$ iterations per day.}
	\label{FigContriARUniformOverview}
\end{figure}

Figure \ref{FigContriARUniformOverview} displays the objective values of the online threshold algorithm for the whole range of percentiles and all factor choices introduced in Section \ref{S3OnlineAlg} using $k=12$ iterations per day. One key observation of Figure \ref{FigContriARUniformOverview} is that the differences in performance between the various factors grow slightly with the percentile. As can be seen, the piece-wise constant factor performed best among all factors (which we also observed for the remaining error distributions and numbers of iterations not displayed in Figure \ref{FigContriARUniformOverview}). Therefore, we restrict our following analysis to the piece-wise constant factor.\\

\subsubsection{Iterations}
\begin{figure}
\centering
\pgfplotsset{every axis legend/.append style={at={(0.5, 1.3)}, anchor = north, legend columns = 3}}
\begin{tikzpicture}
	\begin{axis}[xmin=0, xmax=1, ymin=0, ymax=32,
		xlabel={Percentile},
		ylabel={add. PV $[kWh]$},
		grid = both,
		minor tick num = 1,
		major grid style = {lightgray},
		minor grid style = {lightgray!25},
		legend cell align={left}]
		\addplot[red, thick]{-10};
		\addplot[blue, thick]{-10};
		\addplot[green, thick]{-10};
		\addplot[orange, thick]{-10};
		\addplot[yellow, thick]{-10};
		\addplot[purple, thick]{-10};
		\addplot[black, dashed]{-10};
		\addplot[black, mark = triangle*]{-10};
		\addplot[black, mark = o]{-10};
		\addplot[red, dashed, domain=0:1]{9.558}; 
		\addplot[red, mark = triangle*]table[col sep = comma, x={P}, y={2404}]{FinalContriUniform.csv};
		\addplot[red, mark = o]table[col sep = comma, x={P}, y={2414}]{FinalContriUniform.csv};
		\addplot[blue, dashed, domain=0:1]{13.148};
		\addplot[blue, mark = triangle*]table[col sep = comma, x={P}, y={1604}]{FinalContriUniform.csv};
		\addplot[blue, mark = o]table[col sep = comma, x={P}, y={1614}]{FinalContriUniform.csv};
		\addplot[green, dashed, domain=0:1]{15.894};
		\addplot[green, mark = triangle*]table[col sep = comma, x={P}, y={1204}]{FinalContriUniform.csv};
		\addplot[green, mark = o]table[col sep = comma, x={P}, y={1214}]{FinalContriUniform.csv};
		\addplot[orange, dashed, domain=0:1]{19.703};
		\addplot[orange, mark = triangle*]table[col sep = comma, x={P}, y={804}]{FinalContriUniform.csv};
		\addplot[orange, mark = o]table[col sep = comma, x={P}, y={814}]{FinalContriUniform.csv};
		\addplot[yellow, dashed, domain=0:1]{24.147};
		\addplot[yellow, mark = triangle*]table[col sep = comma, x={P}, y={404}]{FinalContriUniform.csv};
		\addplot[yellow, mark = o]table[col sep = comma, x={P}, y={414}]{FinalContriUniform.csv};
		\addplot[purple, dashed, domain=0:1]{25.956};
		\addplot[purple, mark = triangle*]table[col sep = comma, x={P}, y={204}]{FinalContriUniform.csv};
		\addplot[purple, mark = o]table[col sep = comma, x={P}, y={214}]{FinalContriUniform.csv};
		\legend{iterations 4, iterations 6, iterations 8, iterations 12, iterations 24, iterations 48, OptPath, AR approach, HR apporach}
	\end{axis}
\end{tikzpicture}
\caption{Contributions of uniform online threshold algorithms for the AR and the HR approaches with the piece-wise constant factor and different numbers of iterations (colors) and percentiles (x-axis) for a time horizon of three days}
\label{FigContriUniform}
\end{figure}
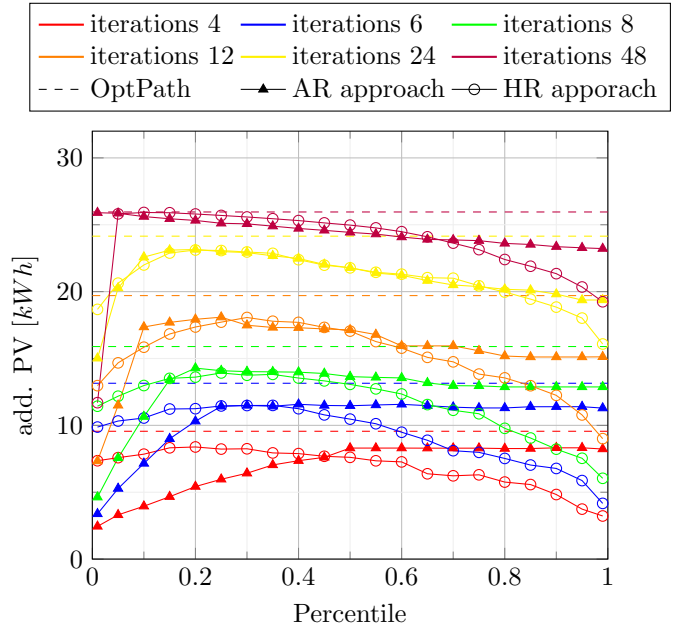

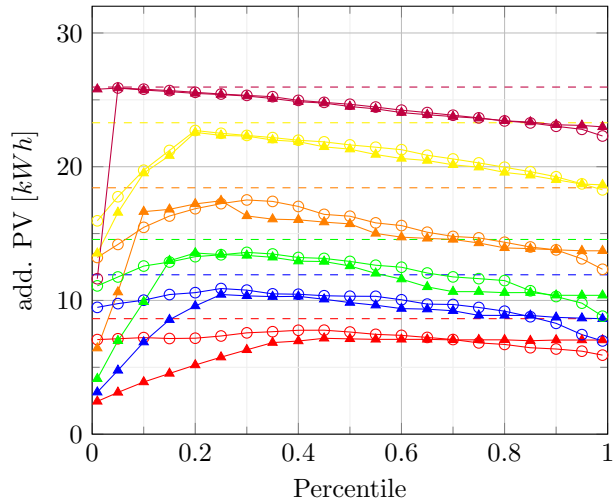
\begin{figure}
\centering
\pgfplotsset{every axis legend/.append style={at={(0.5, 1.5)}, anchor = north, legend columns = 2}}
\begin{tikzpicture}
	\begin{axis}[xmin=0, xmax=1, ymin=0, ymax=32,
		xlabel={Percentile},
		ylabel={add. PV $[kWh]$},
		grid = both,
		minor tick num = 1,
		major grid style = {lightgray},
		minor grid style = {lightgray!25}]
		\addplot[black]{-10};
		\addplot[black, dashed]{-10};
		\addplot[black, mark = triangle*]{-10};
		\addplot[black, mark = o]{-10};
		\addplot[red]{-10};
		\addplot[blue]{-10};
		\addplot[green]{-10};
		\addplot[orange]{-10};
		\addplot[yellow]{-10};
		\addplot[purple]{-10};
		\addplot[red, dashed, domain=0:1]{8.636}; 
		\addplot[red, mark = triangle*]table[col sep = comma, x={P}, y={2404}]{FinalContriNormal.csv};
		\addplot[red, mark = o]table[col sep = comma, x={P}, y={2414}]{FinalContriNormal.csv};
		\addplot[blue, dashed, domain=0:1]{11.918};
		\addplot[blue, mark = triangle*]table[col sep = comma, x={P}, y={1604}]{FinalContriNormal.csv};
		\addplot[blue, mark = o]table[col sep = comma, x={P}, y={1614}]{FinalContriNormal.csv};
		\addplot[green, dashed, domain=0:1]{14.559};
		\addplot[green, mark = triangle*]table[col sep = comma, x={P}, y={1204}]{FinalContriNormal.csv};
		\addplot[green, mark = o]table[col sep = comma, x={P}, y={1214}]{FinalContriNormal.csv};
		\addplot[orange, dashed, domain=0:1]{18.424};
		\addplot[orange, mark = triangle*]table[col sep = comma, x={P}, y={804}]{FinalContriNormal.csv};
		\addplot[orange, mark = o]table[col sep = comma, x={P}, y={814}]{FinalContriNormal.csv};
		\addplot[yellow, dashed, domain=0:1]{23.287};
		\addplot[yellow, mark = triangle*]table[col sep = comma, x={P}, y={404}]{FinalContriNormal.csv};
		\addplot[yellow, mark = o]table[col sep = comma, x={P}, y={414}]{FinalContriNormal.csv};
		\addplot[purple, dashed, domain=0:1]{25.956};
		\addplot[purple, mark = triangle*]table[col sep = comma, x={P}, y={204}]{FinalContriNormal.csv};
		\addplot[purple, mark = o]table[col sep = comma, x={P}, y={214}]{FinalContriNormal.csv};
	\end{axis}
\end{tikzpicture}
\caption{Contributions of normal online threshold algorithms for the AR and the HR approaches with different numbers of iterations (colors) and percentiles (x-axis) for a time horizon of three days. Legend is the same as in Fig. \ref{FigContriUniform}.}
\label{FigContriNormal}
\end{figure}

\begin{figure}
\centering
\pgfplotsset{every axis legend/.append style={at={(0.5, 1.3)}, anchor = north, legend columns = 2}}
\begin{tikzpicture}
	\begin{axis}[xmin=0.5, xmax=6.5, ymin=0, ymax=32,
		xlabel={Iterations},
		ylabel={add. PV $[kWh]$},
		xtick = {1,2,3,4,5,6},
		xticklabels={4,6,8,12,24,48},
		grid = both,
		minor tick num = 1,
		major grid style = {lightgray},
		minor grid style = {lightgray!25},
		legend cell align={left}]
		\addplot[red, dashed, mark = triangle*]{-10};
		\addplot[red, mark = triangle]{-10};
		\addplot[blue, dashed, mark = otimes*]{-10};
		\addplot[blue, mark = o]{-10};
		\addplot[black, dashed, mark = square*]{-10};
		\addplot[black, mark = square]{-10};
		\addplot[red, dashed, mark=triangle*] table {
			1  3.886
			2  6.485
			3  9.442
			4  14.247
			5  21.163
			6  25.942
		};
		\addplot[red, mark=triangle] table {
			1  5.647
			2  8.744
			3  11.066
			4  13.880
			5  20.345
			6  24.753
		};
		\addplot[blue, dashed, mark=o*] table {
			1  3.97
			2  6.559
			3  9.641
			4  14.219
			5  21.143
			6  25.940
		};
		\addplot[blue, mark=o] table {
			1  6.234
			2  9.029
			3  10.904
			4  14.204
			5  19.034
			6  24.696
		};
		\addplot[dashed, mark=square*] table {
			1  4.599
			2  7.573
			3  11.196
			4  16.478
			5  24.491
			6  30.068
		};
		\addplot[ mark=square] table {
			1  6.465
			2  9.538
			3  11.979
			4  15.874
			5  21.831
			6  28.558
		};
		\legend{PR Uniform, exponential \cite{OKPKeywordAuctions2008Zhou} Uniform, PR Normal, exponential \cite{OKPKeywordAuctions2008Zhou} Normal, PR Shifted, exponential \cite{OKPKeywordAuctions2008Zhou} Shifted}
	\end{axis}
\end{tikzpicture}
\caption{Contributions of PR model and exponential \cite{OKPKeywordAuctions2008Zhou} for various numbers of iterations (4 - 48) and uniform, normal, and shifted error distributions}
\label{FigContriPRChakra}
\end{figure}
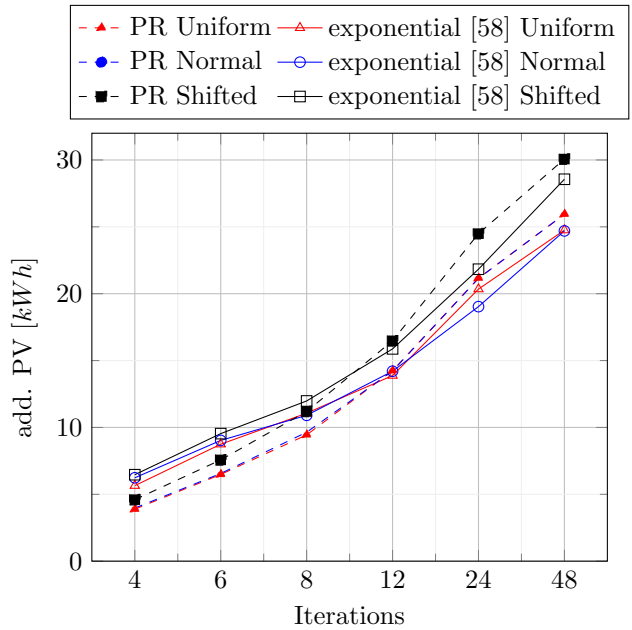

Figures \ref{FigContriUniform} - \ref{FigContriShifted} show the results of the AR and HR approaches for the piece-wise constant factor for the whole range of percentiles and suitably chosen iterations for the three underlying error distributions. The displayed number of maximal iterations per day corresponds to various step sizes of the classical rolling horizon. Figure \ref{FigContriPRChakra} displays the results for the PR and exponential (\cite{OKPKeywordAuctions2008Zhou}) approaches for the same distributions and iterations. We compare the results to the (in hindsight) optimal path solution, denoted by OptPath.\\

\begin{figure}
\centering
\pgfplotsset{every axis legend/.append style={at={(0.5, 1.5)}, anchor = north, legend columns = 2}}
\begin{tikzpicture}
	\begin{axis}[xmin=0, xmax=1, ymin=0, ymax=32,
		xlabel={Percentile},
		ylabel={add. PV $[kWh]$},
		grid = both,
		minor tick num = 1,
		major grid style = {lightgray},
		minor grid style = {lightgray!25}]
		\addplot[black]{-10};
		\addplot[black, dashed]{-10};
		\addplot[black, mark = triangle*]{-10};
		\addplot[black, mark = o]{-10};
		\addplot[red]{-10};
		\addplot[blue]{-10};
		\addplot[green]{-10};
		\addplot[orange]{-10};
		\addplot[yellow]{-10};
		\addplot[purple]{-10};
		\addplot[red, dashed, domain=0:1]{9.85}; 
		\addplot[red, mark = triangle*]table[col sep = comma, x={P}, y={2404}]{FinalContriShifted.csv};
		\addplot[red, mark = o]table[col sep = comma, x={P}, y={2414}]{FinalContriShifted.csv};
		\addplot[blue, dashed, domain=0:1]{13.624};
		\addplot[blue, mark = triangle*]table[col sep = comma, x={P}, y={1604}]{FinalContriShifted.csv};
		\addplot[blue, mark = o]table[col sep = comma, x={P}, y={1614}]{FinalContriShifted.csv};
		\addplot[green, dashed, domain=0:1]{16.692};
		\addplot[green, mark = triangle*]table[col sep = comma, x={P}, y={1204}]{FinalContriShifted.csv};
		\addplot[green, mark = o]table[col sep = comma, x={P}, y={1214}]{FinalContriShifted.csv};
		\addplot[orange, dashed, domain=0:1]{21.163};
		\addplot[orange, mark = triangle*]table[col sep = comma, x={P}, y={804}]{FinalContriShifted.csv};
		\addplot[orange, mark = o]table[col sep = comma, x={P}, y={814}]{FinalContriShifted.csv};
		\addplot[yellow, dashed, domain=0:1]{26.886};
		\addplot[yellow, mark = triangle*]table[col sep = comma, x={P}, y={404}]{FinalContriShifted.csv};
		\addplot[yellow, mark = o]table[col sep = comma, x={P}, y={414}]{FinalContriShifted.csv};
		\addplot[purple, dashed, domain=0:1]{30.087};
		\addplot[purple, mark = triangle*]table[col sep = comma, x={P}, y={204}]{FinalContriShifted.csv};
		\addplot[purple, mark = o]table[col sep = comma, x={P}, y={214}]{FinalContriShifted.csv};
	\end{axis}
\end{tikzpicture}
\caption{Contributions of shifted online threshold algorithms for the AR and the HR approaches with different numbers of iterations (colors) and percentiles (x-axis) for a time horizon of three days, Legend is the same as in Fig. \ref{FigContriUniform}.}
\label{FigContriShifted}
\end{figure}
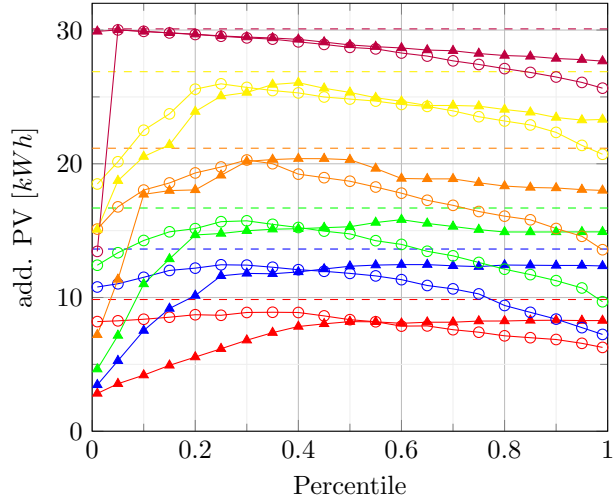

We first analyze the impact of the maximal number of iterations on the solutions. Based on the results observed in \cite{DynRHRMES2023Hoenenetal}, we expect to see an improvement in objective value with an increasing number of iterations. Figures \ref{FigContriUniform} - \ref{FigContriPRChakra} confirm this expectation to a large extent. There are only a few small exceptions to this insight, in particular for extreme percentile choices. Nevertheless, in general, there is a clear distinction in the objective value between the different maximal numbers of iterations, also for the PR approach, see Figure \ref{FigContriPRChakra}. This improvement stems from the additional number of updated PV forecasts with (possibly) improved lower bounds on the PV generation and can be observed for all underlying probability distributions. In addition, it can be observed that the gap towards the optimal solution gets smaller the more iterations can be started.  

\subsubsection{Percentiles}
Focusing on the percentile choice, we notice that for small and large values, the online threshold algorithms perform much worse, see Figures \ref{FigContriUniform} - \ref{FigContriShifted}. This can be explained by the resulting threshold values of the extreme percentile choices. As already mentioned, particularly large or small threshold values can decrease the performance of the overall online algorithm significantly. A balanced percentile choice between 0.15 and 0.5 on the other hand ensures a good performance, although small differences between the underlying probability distributions and approaches can be observed.

\subsubsection{Approaches}
Comparing the results of the AR, HR, and PR approaches with each other, some conclusions can be drawn. First of all, there does not seem to be a large difference between the AR and HR approaches w.r.t. their best objective value. In particular for the normal distribution (see Figure \ref{FigContriNormal}), the two approaches seem to overlap to a high degree for a percentile larger than 0.2, with the best results usually achieved by a percentile choice between 0.2 and 0.3. For the uniform distribution, the results of the AR approach improve with increasing percentile for few iterations, while for the HR approach, it seems to be the opposite. For a larger number of iterations, this observation does not hold any longer, although the AR approach seems to be less sensitive towards large percentile choices compared to the HR approach, which performs better for smaller percentile choices. The best percentile choice once again seems to be within the interval $[0.2,0.3]$ in most cases. For the shifted normal distribution on the other hand the best percentile choice for the AR approach mostly changes to a slightly higher value between 0.4 and 0.6. The only exception is the largest number of iterations, for which a very small percentile choice performs best. Here, the increased percentiles balance out the rather neutral guess of average realizations compared to the distribution mean of 0.4. The HR approach on the other hand is not influenced by the mean of 0.4, as this information is already present in the historical data. \\

In contrast to the AR and HR approaches, the PR approach is independent of the percentile choice. In particular for scenarios with a large number of iterations, it performs quite well, while for few iterations, it is clearly outperformed by the AR and HR approaches.\\

Summarizing, the AR and HR approaches both seem to be sensitive toward the base threshold, while the factor does not influence the outcome a lot. Nevertheless, there are small differences between the factors, and the piece-wise constant factor emerged as the most promising one. The underlying error distributions do not have a large impact on the outcome of the online algorithms. The PR approach performed really well for a larger number of iterations, while some deficits can be observed for a smaller number of iterations.

\subsection{Analysis Online Rolling Horizon}
Within this section, we investigate whether the insights gained in the previous section also transfer to the solutions when taking the whole online rolling horizon-based robust energy management approach into account. In addition to the objective value, which now represents the electricity costs of the microgrid, we also consider and discuss the impact of the online scheme on the local PV usage. We compare the results of the various online algorithms to the solutions of the starting time slots obtained by the optimal offline path solution, denoted by OptPath rolling horizon, as well as the classical and the dynamic rolling horizon versions. We consider various step sizes of the classical rolling horizon and translate these into the corresponding maximal number of iterations for the dynamic and online rolling horizon models, e.g., a step size of 24 time slots results in 4 iterations per day. For the sake of clarity, we did not include the objective values of the classical rolling horizon approach in the figures but still refer to them within the analysis. Similar to the analysis presented in Section \ref{S52AOTA}, we restrict the analysis to the percentile range $\left[ 0.15, 0.5 \right]$ as well as the piece-wise constant factor.\\

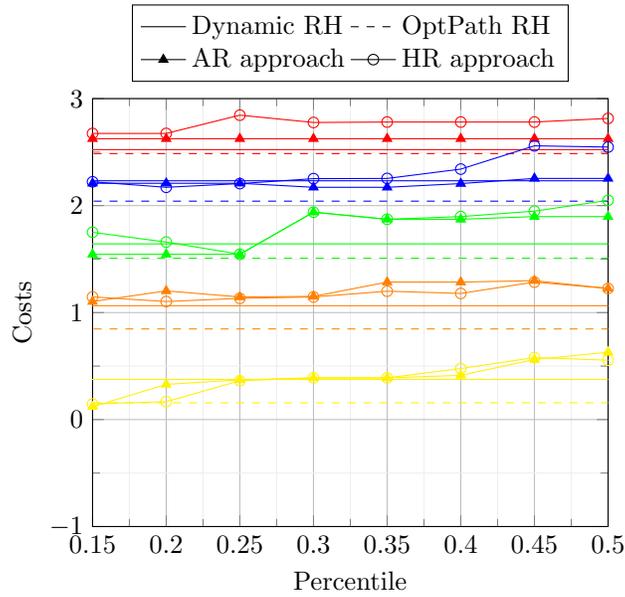
\begin{figure}
\centering
\pgfplotsset{every axis legend/.append style={at={(0.5, 1.22)}, anchor = north, legend columns = 2}}
\begin{tikzpicture}
	\begin{axis}[xmin=0.15, xmax=0.5, ymin=-1, ymax=3,
		xlabel={Percentile},
		ylabel={Costs},
		grid = both,
		minor tick num = 1,
		major grid style = {lightgray},
		minor grid style = {lightgray!25},
		legend cell align={left}]
		\addplot[black]{-10};
		\addplot[black, dashed]{-10};
		\addplot[black, mark=triangle*]{-10};
		\addplot[black, mark=o]{-10};
		\addplot[red, domain=0:0.5]{2.523}; 
		\addplot[red, dashed, domain=0:0.5]{2.4858};
		\addplot[red, mark = triangle*]table[col sep = comma, x={P}, y={2404}]{FinalCostUniform.csv};
		\addplot[red, mark = o]table[col sep = comma, x={P}, y={2414}]{FinalCostUniform.csv};
		\addplot[blue, domain=0:0.5]{2.232};
		\addplot[blue, dashed, domain=0:0.5]{2.0416};
		\addplot[blue, mark = triangle*]table[col sep = comma, x={P}, y={1604}]{FinalCostUniform.csv};
		\addplot[blue, mark = o]table[col sep = comma, x={P}, y={1614}]{FinalCostUniform.csv};
		\addplot[green, domain=0:0.5]{1.642};
		\addplot[green, dashed, domain=0:0.5]{1.508};
		\addplot[green, mark = triangle*]table[col sep = comma, x={P}, y={1204}]{FinalCostUniform.csv};
		\addplot[green, mark = o]table[col sep = comma, x={P}, y={1214}]{FinalCostUniform.csv};
		\addplot[orange, domain=0:0.5]{1.064};
		\addplot[orange, dashed, domain=0:0.5]{0.847};
		\addplot[orange, mark = triangle*]table[col sep = comma, x={P}, y={804}]{FinalCostUniform.csv};
		\addplot[orange, mark = o]table[col sep = comma, x={P}, y={814}]{FinalCostUniform.csv};
		\addplot[yellow, domain=0:0.5]{0.375};
		\addplot[yellow, dashed, domain=0:0.5]{0.157};
		\addplot[yellow, mark = triangle*]table[col sep = comma, x={P}, y={404}]{FinalCostUniform.csv};
		\addplot[yellow, mark = o]table[col sep = comma, x={P}, y={414}]{FinalCostUniform.csv};
		\legend{Dynamic RH, OptPath RH, AR approach, HR approach}
	\end{axis}
\end{tikzpicture}
\caption{Objective value (in costs) of online algorithms for various numbers of iterations (colors) and percentiles (x-axis) for a time horizon of three days with uniform uncertainty distribution. The color scheme is the same as used before (see Fig. \ref{FigContriUniform}).}
\label{FigCostsUniform}
\end{figure}

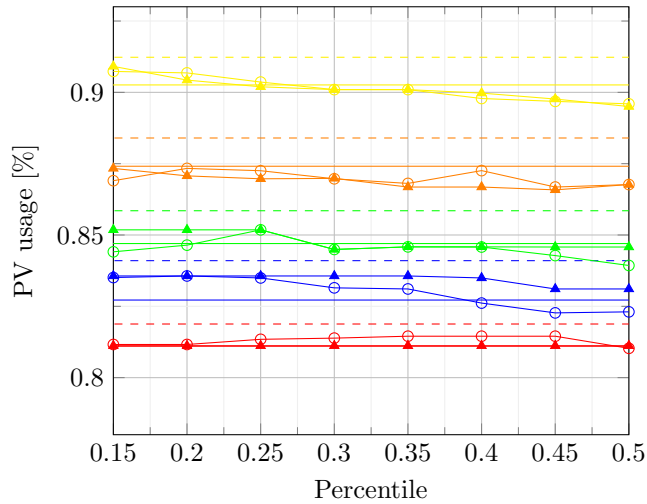
\begin{figure}
\centering
\pgfplotsset{every axis legend/.append style={at={(0.5, 1.215)}, anchor = north, legend columns = 2}}
\begin{tikzpicture}
	\begin{axis}[xmin=0.15, xmax=0.5, ymin=0.78, ymax=0.93,
		xlabel={Percentile},
		ylabel={PV usage [\%]},
		grid = both,
		minor tick num = 1,
		major grid style = {lightgray},
		minor grid style = {lightgray!25}]
		\addplot[black]{-1};
		\addplot[black, dashed]{-1};
		\addplot[black, mark=triangle*]{-1};
		\addplot[black, mark=o]{-1};
		\addplot[red, domain=0:0.5]{0.81103};
		\addplot[red, dashed, domain=0:0.5]{0.8188};
		\addplot[red, mark = triangle*]table[col sep = comma, x={P}, y={2404}]{FinalPVUniform.csv};
		\addplot[red, mark = o]table[col sep = comma, x={P}, y={2414}]{FinalPVUniform.csv};
		\addplot[blue, domain=0:0.5]{0.8272};
		\addplot[blue, dashed, domain=0:0.5]{0.8410};
		\addplot[blue, mark = triangle*]table[col sep = comma, x={P}, y={1604}]{FinalPVUniform.csv};
		\addplot[blue, mark = o]table[col sep = comma, x={P}, y={1614}]{FinalPVUniform.csv};
		\addplot[green, domain=0:0.5]{0.8470};
		\addplot[green, dashed, domain=0:0.5]{0.8585}; 
		\addplot[green, mark = triangle*]table[col sep = comma, x={P}, y={1204}]{FinalPVUniform.csv};
		\addplot[green, mark = o]table[col sep = comma, x={P}, y={1214}]{FinalPVUniform.csv};
		\addplot[orange, domain=0:0.5]{0.8741};
		\addplot[orange, dashed, domain=0:0.5]{0.8840};
		\addplot[orange, mark = triangle*]table[col sep = comma, x={P}, y={804}]{FinalPVUniform.csv};
		\addplot[orange, mark = o]table[col sep = comma, x={P}, y={814}]{FinalPVUniform.csv};
		\addplot[yellow, domain=0:0.5]{0.9026};
		\addplot[yellow, dashed, domain=0:0.5]{0.9123}; 
		\addplot[yellow, mark = triangle*]table[col sep = comma, x={P}, y={404}]{FinalPVUniform.csv};
		\addplot[yellow, mark = o]table[col sep = comma, x={P}, y={414}]{FinalPVUniform.csv};
	\end{axis}
\end{tikzpicture}
\caption{PV usage (in \%) of online algorithms for various numbers of iterations (colors) and percentiles (x-axis) for a time horizon of three days with uniform uncertainty distribution. The color scheme is the same as used before (see Fig. \ref{FigContriUniform}).}
\label{FigPVUniform}
\end{figure}

We start with the results based on uniformly random realizations. Figure \ref{FigCostsUniform} shows the objective function values, which represent the electricity costs of the microgrid, while Figure \ref{FigPVUniform} shows the local PV usage of the solutions. Note that we did not include the results for the PR or the exponential approach in the figures, as they usually performed significantly worse than the displayed approaches. Therefore, we focus our analysis on the AR and HR approaches:
\begin{itemize}
\item At first glance, when directly comparing Figures \ref{FigCostsUniform} and \ref{FigPVUniform}, we already notice that the results regarding objective function and local PV usage rate seem to mirror each other to a high level of detail. This once again highlights the impact of the PV forecasts and generation on the objective value and strengthens our choice of using the PV forecasts as the underlying measure for the online algorithms within this application scenario.
\item In general, the insights gained during the analysis of the online threshold algorithms also transfer to the results including the robust energy management scheme. The number of iterations has a significant impact on the solution, both regarding costs and local PV usage, and the more often an iteration can be started, the better the results. The individual percentile choices also align well with the previous results and in general, there is no clear difference between the AR and HR approaches, neither w.r.t. the objective values nor regarding the PV usage. Both perform similarly when considering the percentiles and iterations.
\item Compared to the dynamic rolling horizon solution, the AR and the HR approach outperform it for suitably chosen percentiles both, in costs, as well as in local PV usage. The online algorithms achieve results that are up to 50\% better compared to the dynamic rolling horizon results, see Figure \ref{FigCostsUniform}. The results of the classical rolling horizon model are not displayed in the figures, as the objective value is much larger compared to all other approaches, with improvements of the online algorithms of over 85\% compared to the classical rolling horizon. This also aligns well with the insights from \cite{DynRHRMES2023Hoenenetal}, in which the dynamic rolling horizon also clearly outperformed the classical version.
\item Comparing the AR and HR approaches against the OptPath rolling horizon, which is based on the optimal offline starting time slots, as presented in the previous analysis, we notice that the more iterations can be started, the better the online algorithms perform. In a few cases, the objective values of the online approaches are even better than the ones of the optimal starting time slot rolling horizon model. At first glance, this may seem counterintuitive, as the OptPath model served as a bound of the optimal solution. However, the online threshold algorithms are solely based on the PV forecasts, while the underlying robust energy management problem also considers uncertainties in market prices, or EV and household load. Hence, different decisions taken during the robust energy management approach combined with uncertainty in market prices have led to a situation, in which the optimal starting time slots performed worse than the online starting time slots. When focusing on the PV usage (see Figure \ref{FigPVUniform}), which better represents the objective of the online threshold algorithms, the optimal starting time slot rolling horizon model clearly outperforms the online algorithms again.
\end{itemize}

\begin{figure}
\centering
\pgfplotsset{every axis legend/.append style={at={(0.5, 1.125)}, anchor = north, legend columns = 2}}
\begin{tikzpicture}
	\begin{axis}[xmin=0.15, xmax=0.5, ymin=-1, ymax=3,
		xlabel={Percentile},
		ylabel={Costs},
		grid = both,
		minor tick num = 1,
		major grid style = {lightgray},
		minor grid style = {lightgray!25}]
		\addplot[black]{-10};
		\addplot[black, dashed]{-10};
		\addplot[black, mark=triangle*]{-10};
		\addplot[black, mark=o]{-10};
		\addplot[red, domain=0:0.5]{2.5906}; 
		\addplot[red, dashed, domain=0:0.5]{2.6233}; 
		\addplot[red, mark = triangle*]table[col sep = comma, x={P}, y={2404}]{FinalCostNormal.csv};
		\addplot[red, mark = o]table[col sep = comma, x={P}, y={2414}]{FinalCostNormal.csv};
		\addplot[blue, domain=0:0.5]{2.254};
		\addplot[blue, dashed, domain=0:0.5]{2.196};
		\addplot[blue, mark = triangle*]table[col sep = comma, x={P}, y={1604}]{FinalCostNormal.csv};
		\addplot[blue, mark = o]table[col sep = comma, x={P}, y={1614}]{FinalCostNormal.csv};
		\addplot[green, domain=0:0.5]{1.895};
		\addplot[green, dashed, domain=0:0.5]{1.7387};
		\addplot[green, mark = triangle*]table[col sep = comma, x={P}, y={1204}]{FinalCostNormal.csv};
		\addplot[green, mark = o]table[col sep = comma, x={P}, y={1214}]{FinalCostNormal.csv};
		\addplot[orange, domain=0:0.5]{1.094};
		\addplot[orange, dashed, domain=0:0.5]{1.137};
		\addplot[orange, mark = triangle*]table[col sep = comma, x={P}, y={804}]{FinalCostNormal.csv};
		\addplot[orange, mark = o]table[col sep = comma, x={P}, y={814}]{FinalCostNormal.csv};
		\addplot[yellow, domain=0:0.5]{0.354};
		\addplot[yellow, dashed, domain=0:0.5]{0.2917};
		\addplot[yellow, mark = triangle*]table[col sep = comma, x={P}, y={404}]{FinalCostNormal.csv};
		\addplot[yellow, mark = o]table[col sep = comma, x={P}, y={414}]{FinalCostNormal.csv};
	\end{axis}
\end{tikzpicture}
\caption{Objective value (in costs) of online algorithms for various numbers of iterations (colors) and percentiles (x-axis) for a time horizon of three days with normal uncertainty distribution. The color scheme is the same as used before (see Fig. \ref{FigContriUniform}).}
\label{FigCostsNormal}
\end{figure}
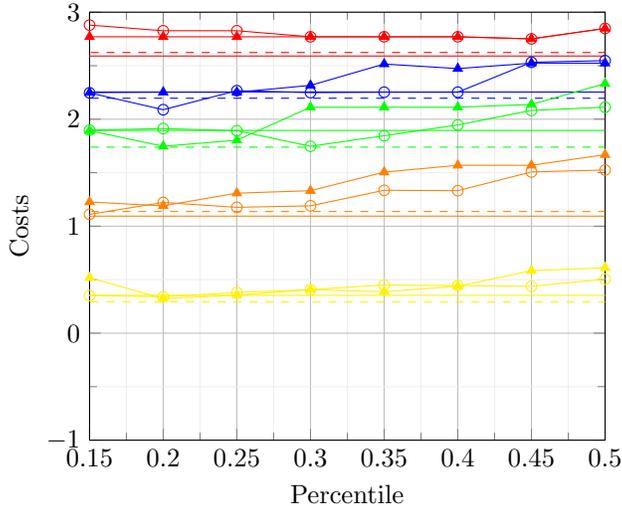

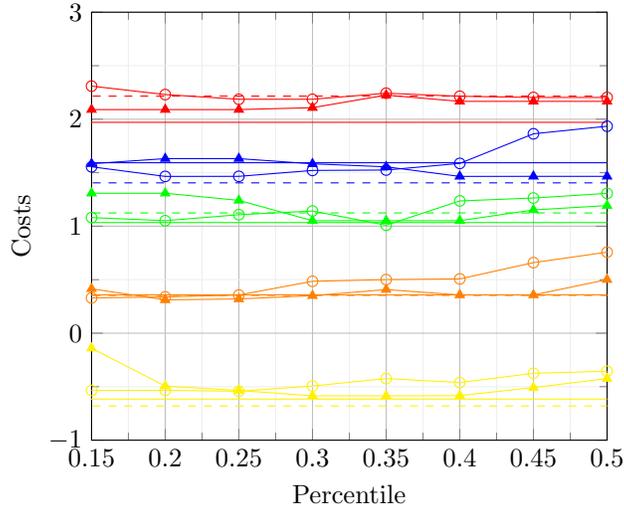
\begin{figure}
\centering
\pgfplotsset{every axis legend/.append style={at={(0.5, 1.125)}, anchor = north, legend columns = 2}}
\begin{tikzpicture}
	\begin{axis}[xmin=0.15, xmax=0.5, ymin=-1, ymax=3,
		xlabel={Percentile},
		ylabel={Costs},
		grid = both,
		minor tick num = 1,
		major grid style = {lightgray},
		minor grid style = {lightgray!25}]
		\addplot[black]{-1.5};
		\addplot[black, dashed]{-1.5};
		\addplot[black, mark=triangle*]{-1.5};
		\addplot[black, mark=o]{-1.5};
		\addplot[red, domain=0:0.5]{1.971}; 
		\addplot[red, dashed, domain=0:0.5]{2.216}; 
		\addplot[red, mark = triangle*]table[col sep = comma, x={P}, y={2404}]{FinalCostShifted.csv};
		\addplot[red, mark = o]table[col sep = comma, x={P}, y={2414}]{FinalCostShifted.csv};
		\addplot[blue, domain=0:0.5]{1.593};
		\addplot[blue, dashed, domain=0:0.5]{1.405};
		\addplot[blue, mark = triangle*]table[col sep = comma, x={P}, y={1604}]{FinalCostShifted.csv};
		\addplot[blue, mark = o]table[col sep = comma, x={P}, y={1614}]{FinalCostShifted.csv};
		\addplot[green, domain=0:0.5]{1.034};
		\addplot[green, dashed, domain=0:0.5]{1.124};
		\addplot[green, mark = triangle*]table[col sep = comma, x={P}, y={1204}]{FinalCostShifted.csv};
		\addplot[green, mark = o]table[col sep = comma, x={P}, y={1214}]{FinalCostShifted.csv};
		\addplot[orange, domain=0:0.5]{0.3583};
		\addplot[orange, dashed, domain=0:0.5]{0.353};
		\addplot[orange, mark = triangle*]table[col sep = comma, x={P}, y={804}]{FinalCostShifted.csv};
		\addplot[orange, mark = o]table[col sep = comma, x={P}, y={814}]{FinalCostShifted.csv};
		\addplot[yellow, domain=0:0.5]{-0.6157};
		\addplot[yellow, dashed, domain=0:0.5]{-0.6803};
		\addplot[yellow, mark = triangle*]table[col sep = comma, x={P}, y={404}]{FinalCostShifted.csv};
		\addplot[yellow, mark = o]table[col sep = comma, x={P}, y={414}]{FinalCostShifted.csv};
	\end{axis}
\end{tikzpicture}
\caption{Objective value (in costs) of online algorithms for various numbers of iterations (colors) and percentiles (x-axis) for a time horizon of three days with shifted normal uncertainty distribution. The color scheme is the same as used before (see Fig. \ref{FigContriUniform}).}
\label{FigCostsShifted}
\end{figure}

The results for the normal (Figure \ref{FigCostsNormal}), respectively shifted normal distribution (Figure \ref{FigCostsShifted}) slightly differ from the uniform results. Nevertheless, the general insights from the analysis of the online scheduling algorithms, such as the impact of the number of iterations or the percentile choice on the objective value or the similarity of the AR and the HR approaches, are still valid. The main differences to the uniform uncertainty realizations are the results of the dynamic rolling horizon framework, which are much better and often outperform the OptPath approach. Nevertheless, for selected iterations and percentiles, the online algorithms still achieve better results than the dynamic or OptPath approach (see Figures \ref{FigCostsNormal} and \ref{FigCostsShifted}), although the improvements are much smaller compared to the uniform distribution case. The classical rolling horizon results are once again not displayed as these perform significantly worse than any of the other considered rolling horizon schemes, with improvements of the online algorithms of over 80\%.

\subsection{Discussion}
Summarizing the previous analysis of the results, we notice that the online algorithms clearly outperform the dynamic rolling horizon model for the uniform distribution, while the results for the normal and shifted normal distribution are much more similar to each other. The OptPath approach, which served as an upper bound for the pure online algorithms without the energy management approach, sometimes achieves worse results than the online or the dynamic rolling horizon approaches. This can be explained by the additional uncertainty, such as e.g., uncertain market prices, which are not considered in the online threshold algorithm. The differences in performance between the underlying error distributions hint that the online algorithms are particularly useful in certain scenarios. Within the context of the energy management approach, the uniform error distribution could be seen as the representation of a particularly cloudy, but still sunny day, in which PV forecasts can change drastically within a very short time. Hence, both online algorithms are able to deal with such fluctuating forecasts much better than any offline algorithm, highlighting interesting application scenarios.\\

Another interesting insight from the analysis concerns the similarity between the AR and the HR approach. Even though the HR approach is based on much more detailed information in the form of multiple data sets following the same underlying unknown error distribution, the AR approach performed equally well, even if the mean of the underlying error distribution does not align with the zero mean assumption of the AR approach, as can be observed for the shifted normal distribution. Thus, we are able to design well-performing online algorithms using only very little additional information.

\section{Conclusion}\label{S6Conclusion}

The aim of this work was to develop the idea of a tailor-made scheduling scheme for the starting times of iterations of a rolling horizon even further. Given the often highly dynamic nature of real-world problems with ever-changing forecasts, scheduling the iterations of the rolling horizon in an online fashion seems to be a promising idea. Therefore, we developed an online scheduling framework for a rolling horizon, inspired by results and insights from various online optimization problems. We proposed, tested, and analyzed several online threshold algorithms in a case study concerning an energy management problem under uncertainty. The results of the online algorithms meet the expectations and we observed improvements of up to 50\% compared to the offline tailor-made scheduling scheme introduced in \cite{DynRHRMES2023Hoenenetal}. When only focusing on the online algorithms without the underlying robust energy management approach, we were able to achieve optimality gaps of less than 2.7\%. We also highlighted the importance of a well-chosen threshold and the effect it can have on the objective value. One key finding is that we do not need much information on future realizations and uncertainty to design well-performing, simple online algorithms, which are comparable to approaches that rely on far more information regarding error distributions. Furthermore, the online algorithms perform particularly well compared to the offline approaches in situations with highly fluctuating uncertainty realizations. These are situations, in which offline approaches tend to perform badly, emphasizing the importance of an online decision tool to regularly achieve good results.\\

Based on the results and discussion, several interesting future research directions arise. In a first step, a more theoretical sensitivity analysis of the threshold could produce valuable insights for the performance of the online algorithms. In particular, the connection between the percentiles and the underlying error distribution could provide valuable insights, which could be used to design threshold functions, yielding some form of guarantee. A more applied research direction would be to combine the offline and online algorithms into one algorithm, which depending on the realizations can switch between a predetermined iteration schedule and the online decision-making scheme. Such a combined approach can combine the best of both worlds, with the insurance to use all possible iterations due to the offline schedule and the ability to react to unusually good or bad realizations due to the online decision-making.

\section*{Acknowledgements}
This research is supported by the Netherlands Organization for Scientific Research (NWO) Grant 645.002.001.

\printbibliography

\end{document}